\documentclass[12pt,english]{article}

\usepackage[T1]{fontenc}
\usepackage[a4paper]{geometry}
\usepackage{color}
\usepackage{float}
\usepackage{amsmath}
\usepackage{amsthm}
\usepackage{graphicx}
\usepackage{setspace}
\makeatletter

\newcommand{\lyxmathsym}[1]{\ifmmode\begingroup\def\b@ld{bold}
  \text{\ifx\math@version\b@ld\bfseries\fi#1}\endgroup\else#1\fi}

\providecommand{\tabularnewline}{\\}

\newcommand{\lyxaddress}[1]{
\par {\raggedright #1
\vspace{1.4em}
\noindent\par}
}

\date{}

\makeatother

\usepackage{babel}
\begin{document}

\title{In-situ soil parametrization from multi-layer moisture data}

\author{Vitali G.$^{1*}$, Iotti M.$^{2}$, Zambonelli A.$^{1}$}
\maketitle

\lyxaddress{{\footnotesize{}1-University of Bologna, viale Fanin,44, 40127 Bologna,
Italy }\\
{\footnotesize{}2-University of L'Aquila, via Vetoio, 67100 Coppito,
L'Aquila, Italy}\\
{\footnotesize{}{*} corresponding author: giuliano.vitali@unibo.it }}
\begin{abstract}
Inversion methodology has been used to obtain, from multi-layer soil
probes records, a complete soil parametrisation, namely water retention
curve, unsaturated conductivity curve and bulk density at 4 depths.
The approach integrates water dynamics, hysteresis and the effect
of bulk density on conductivity to extract soil parameters required
from most simulation models. The method is applied to sub-sets of
data collection, allowing to understand that not every data-sets contains
the information required for method convergence. A comparison with
experimental bulk-density values show that inversion could give information
even with a better adherence to model, as it considers the effect
of roots and skeleton. The method may be applied to any type of multi-layer
water content probes giving the opportunity to enrich soil parameter
availability and reliability.
\end{abstract}
\begin{quotation}
\textbf{keywords }- inverse problem, soil hydrology, soil structure
\end{quotation}

\section*{Introduction}

Hydrological characterization of soils is a routine laboratory activity,
and parameter values seem to suffice the needs of most soil-based
model used in hydrology, agro-forestry, ecology, etc., whose complexity
induces users to adopt a simplistic view of soil, referencing to standard
soils or pedofunctions, in the belief of a low sensitivity on soil
parametrization.

\textcolor{black}{Soil is a complex system as well and, though the
major lines of its hydrological behavior have been drawn, there are
features not fully captured from math formalism, which are so forth
not jet included in modeling.}

\textcolor{black}{In a porous system, water dynamics is ruled by Darcy-Buckingham
law, where the driving variable is soil water potential (SWP,$\psi$).
S}uch a variable, \textcolor{black}{fundamental in controlling organism
accessibility to water,} is more difficult to measure than Soil Water
Content (SWC,$\theta$). This is the reason why it is fundamental
to know the relation between $\theta$ and $\psi$, the Water Retention
Curve (WRC), long investigated in the domain of hydrology and soil
physics. WRC has been interpreted from a wide series of functions
\cite{Wijaya-2016}, which often fail to represent the two faces of
a soil, micro-and macro-porosity, the latter being related to structure
(aggregation), which in turn depends on clay and organic matter content.

Sampling a soil always means altering its structure. Most of WRCs
are obtained in laboratory by a drying process, generating parameters
that hardly represent the original system, also explaining why soil
hydrology models are hard to be calibrated (\cite{Duan-1992}).

These are the reasons why in-situ soil parametrization represents
a fundamental task. Unfortunately in-situ methods to evaluate physical
and hydrological parameters are complex, time expensive and with large
errors\cite{Durner-2005} therefore a growing number of research have
been oriented to inverse methods (\cite{Vrugt-2008}), aimed at obtaining
the values of the parameters of a model the investigator nesting solution
in a ``non-linear fitting'' methodology (\cite{Tarantola-2005}).

The objective of the present study is to identify parameters of constitutive
relations (WRC and UWC) inverting a soil water dynamics model including
bulk density variability along depth, and dependence of saturated
conductivity on bulk density (see e.g. \cite{Shaykewitch-1970}).

Methodology has been developed to be fed by any soil multi-layer probes
(recording SWC or surrogate variable at different depths), which allows
to collect a long history of data.

In this paper we first describe the experimental setup from which
the data has been described, the model used to interpret collected
data, the inversion methodologies, the parameters obtained and a comparison
with experimental bulk density values.

\section*{Materials and Methods}

\subsection*{Data Records}

SWC records come from an experimental site used for a study on hypogean
fungi (truffle) near Bologna (Italy, location Saiarino, Lat.$44{^\circ}37\lyxmathsym{\textquoteright}N$,
Lon.$11{^\circ}49\lyxmathsym{\textquoteright}E$, $5m$ asl), with
a mediterranean sub-humid climate (sub-continental temperate, after
Koppen) with c.ca $700mm$ precipitation, and mean air temperature
$13{^\circ}C$. The site is located in the basin of Po valley, with
an alluvial soil classified as aquic ustochrept, coarse loamy, mixed,
thermic (USDA Soil Taxonomy - other parameters are available at http://geo.regione.emilia-romagna.it/cartpedo/).
The site is on the banks of a land-reclamation channel, where different
natural contexts may be found. In prior investigations 4 plots (P1..P4)
have been chosen, and found to have slightly different physical parameters
(Table \ref{tab:1}), \cite{Iotti-2018}. 

\begin{table}[h]
\begin{centering}
\begin{tabular}{cccc}
\hline 
plot & \multicolumn{2}{c}{texture} & BD\tabularnewline
\hline 
 & Sand $%\ensuremath{\%}
$ & Clay $%\ensuremath{\%}
$ & $g/cm^{3}$\tabularnewline
\hline 
P1 & 50-60 & < 12 & 1.15-1.25\tabularnewline
\hline 
P2 & 30-40 & 18-25 & 1.45-1.60\tabularnewline
\hline 
P3 & 30-40 & 18-25 & 1.45-1.60\tabularnewline
\hline 
P4 & 30-40 & 18-25 & 1.00-1.15\tabularnewline
\hline 
\end{tabular}
\par\end{centering}
\caption{\label{tab:1}Physical characterization of the experimental plots
from former observations.}
\end{table}

The 4 plots have been delimited and split to have a rain-fed and an\textbf{
}irrigated treatment: water has been supplied on July and August at
intervals of 14 days in 2012 and 7 days in 2013, with 20 mm of water
(dry-weeks only).

DM400 probes (by DFM Software - ZA, \cite{DFM-2010,Mjanyelwa-2016},
equipped with $4$ soil temperature and electric capacity sensors
recording data at the depth of 10, 20, 30 and 40 cm, have been set
on every irrigated sub-plots, and on rain-fed sub-plots of P1,P3.
Hourly data have been recorded from spring 2012 to fall 2013.

Though the probes already return pre-calibrated SWC values, referred
to as $\theta_{DFM}$, they have been re-calibrated in laboratory
so as to correct SWCs by Soil Temperature ($Ts$) and bulk density
($\rho_{aps}$): 
\begin{equation}
\theta=f(\theta_{DFM},T_{s},\rho_{aps})
\end{equation}
 Calibration procedure and relative results are described in Appendix
A.

\subsection*{The Model}

Soil water dynamics around probes is assumed to be 1-D so that the
Darcy-Buckingham can be written in the form: 
\begin{equation}
q=K(\theta)\cdot(1-\partial_{z}\psi_{t})
\end{equation}
 where $q$ is water flux, $K$ the water conductivity, and $\psi_{t}$
the total water potential ($\psi_{t}=\psi_{g}+\psi_{m}=z+\psi_{m}$
, being $\psi_{g}$ the gravitational potential, and $\psi_{m}$ the
matrix potential). The equation can be solved combining it to the
mass-conservation law, $\theta_{t}=q_{z}$ (generating the Richard's
law), defining proper Initial and Boundary Conditions, and adopting
valid expressions for the Water Retention Curve (WRC ,$\psi(\theta)$
), and for the Unsaturated Conductivity Curve (UCC $K(\theta)$ ).
In this study the three-parameter van Genuchten (\cite{VanGenuchten-1980})
function is adopted:

\begin{equation}
\psi=\frac{(S^{n/(1-n)}-1)^{(1/n)}}{a}\quad;\quad0\le S\le1\quad,\quad n>1\quad,\quad a>0
\end{equation}
where $n,a$ are two shape factors, while $S$ is saturation: $S=(\theta\text{\textendash}\theta_{R})/(\theta_{S}\text{\textendash}\theta_{R})$,
where $\theta_{R}$ is residual water content and $\theta_{S}$ is
the saturated SWC: $\theta_{S}=1\lyxmathsym{\textendash}\rho_{aps}/\rho_{S}$
($\rho_{aps}$ being the bulk density and $\rho_{S}$ the real density,
for non-organic soils assumed approximately $2.7g/cm^{3}$).

The model also includes hysteresis, which was accounted following
the approach described in \cite{Elmaloglou-2008} where drying and
wetting WRCs are respectively characterized by the shape factors $a_{d}$
and $a_{w}$ with $a_{d\:}\le\:a_{w}$ . 

The UCC is the one obtained coupling van Genuchten\textquoteright s
WRC to Mualem\textquoteright s model (\cite{Mualem-1976}) :
\begin{equation}
K=K_{S}\cdot S^{b}\cdot[1-(1-S^{1/m})^{m}]^{2};
\end{equation}

commonly adopted with $m=1-1/n$, where $b$ is a shape factor. 

To include the effect of soil structure on saturated conductivity,
$K_{S}$ is assumed to be linearly related to bulk density, as suggested
by results of \cite{Shaykewitch-1970} and \cite{Osunbitan-2005}:

\begin{equation}
K_{S}=K_{\mu}+c\cdot(\rho_{\mu}-\rho_{aps})\quad,\quad\rho_{\mu}\succeq\rho_{aps}
\end{equation}

where $K_{\mu}$ is the value of saturated conductivity for the compact
soil (maximum bulk density:$\rho_{\mu}$), and $c$ is a proportionality
coefficient depending on soil compaction.

\begin{figure}[h]
\begin{centering}
\begin{tabular}{cc}
\includegraphics[width=8cm]{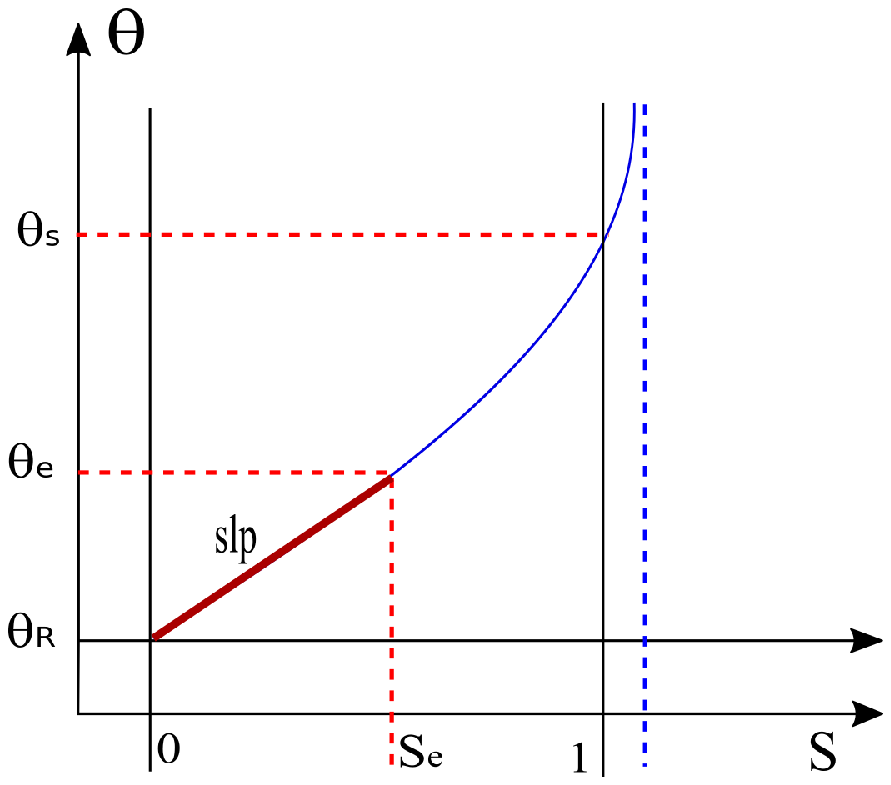} & \includegraphics[width=8cm]{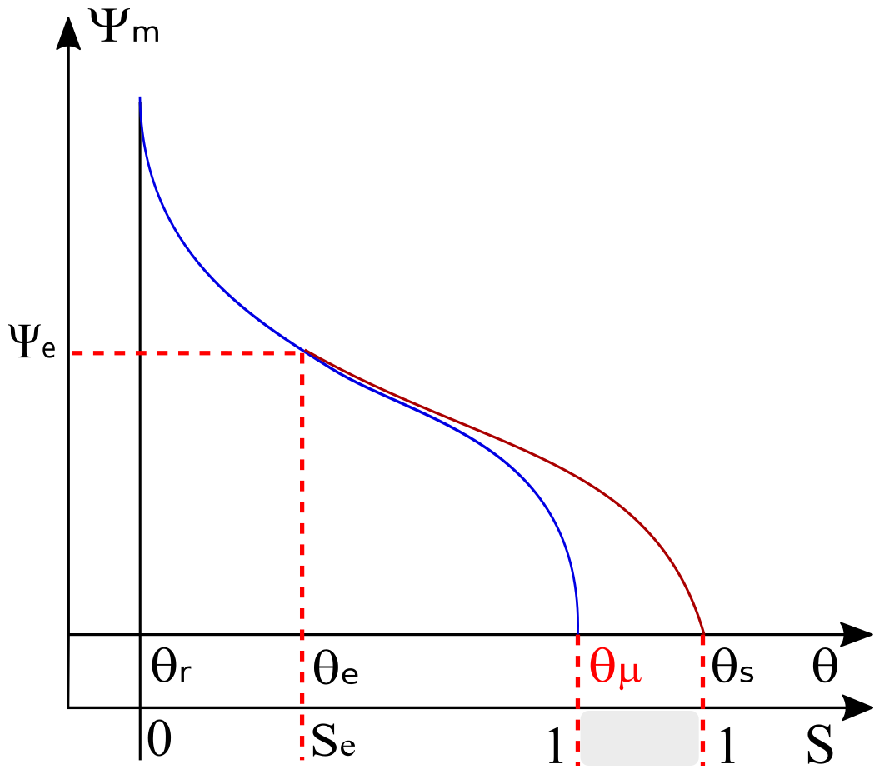}\tabularnewline
\end{tabular}\caption{\label{fig:1}Stretching scheme adopted to include macro-porosity
in WRC}
\par\end{centering}
\end{figure}

To include macro-porosity effect on WRC, instead of considering bi-modality,
we adopt a stretching technique. Starting from the air-entry potential,
$\psi_{e}=1/a$, a corresponding SWC value can be obtained $\theta_{e}=(\theta_{\mu}-\theta_{R})\cdot S_{e}+\theta_{R}$
, being a saturation value $S_{e}=S(\psi_{e},a,n)=2^{(1-n)/n}$, and
where $\theta_{\mu}$ is the saturation SWC for a compact soil: $\theta_{\mu}=(1-\rho_{\mu}/\rho_{S})$.
So forth we can define the slope that relates SWC and saturation in
the low-saturation branch of WRC, $slp=(\theta_{e}-\theta_{R})/S_{e}$,
from which a trend for the higher SWC values is derived: $\theta=\theta_{R}+slp\cdot erf^{-1}[S\cdot A]$,
where the coefficient $A=erf[(\theta_{S}-\theta_{R})/slp]$ is used
to force $\theta=\theta_{S}$ at $S=1$. The stretching scheme is
represented in Figure \ref{fig:1}.

We finally assume that both WRC and UCC have the same parameters for
the whole profile, while bulk density is changing along depth. Calibration
function is also included in inversion procedure.

\subsection*{Inversion scheme}

The solution scheme (see Appendix B) has been integrated into a standard
parameter fitting procedure to compute expected value of SWC $\bar{\theta}$
at $20cm$ and $30cm$ depth. Though a huge number of collected data
(> 10,000 per probe), the procedure requires that a change in $\theta$
values occurs simultaneously at every depth:
\begin{center}
\textbf{rule-a:} $\theta_{DFM}(t,z)\neq\theta_{DFM}(t+\delta t,z)\:;\:z=10,20,30,40$.
\par\end{center}

a constraint which reduces considerably the number of valid data.
Moreover, \cite{Nelder-1965} suggests that to ensure that data contains
the required information, their size should be sensitively greater
than the number of parameters: $n_{data}\geq2\cdot n_{parameters}+1$,
therefore fitting has been operated on subsets with \textbf{$n_{data}=30$}
valid data.

The fitting techniques adopted in this study owns to a diffused class
of inversion techniques based on Jacobian matrix: the Trust-Region-Reflective
Algorithm \cite{Conn-2000} which is similar to the well known Levenberg-Marquard
method, but it also includes parameter ranges. Minimum and maximum
values of parameters are reported in table \ref{tab:2}, together
with the initial value.

\begin{table}[h]
\begin{centering}
\begin{tabular}{ccccccccc}
\hline 
 & \textbf{$\rho_{aps}$} & \textbf{$\theta_{R}$} & \textbf{$a_{d}$} & \textbf{$a_{w}$} & \textbf{$n$} & \textbf{$K\mu$} & $c$ & \textbf{$b$}\tabularnewline
\hline 
min & 0.9 & 0 & 0.01 & 0.01 & 1 & $10^{-4}$ & 0.3 & 5\tabularnewline
\hline 
max & 1.8 & 0.1 & 0.05 & 0.25 & 4 & $100$ & 3.0 & 20\tabularnewline
\hline 
ini & 1.3 & 0.01 & 0.02 & 0.03 & 2 & $1.0$ & 1.0 & 10\tabularnewline
\hline 
\end{tabular}
\par\end{centering}
\caption{\label{tab:2}Minimum, maximum and initial value of parameters used
in inversion procedures}
\end{table}

Methods of this class are quite sensitive to initial parameter guess
(\cite{Tarantola-2005}), making the algorithm to be trapped in regions
of parameter space with local minims. Nonetheless it is also true
that data selected do not guarantee to own the information required
to parameter identification: even models with few parameters could
reach a complexity making the parameter search prohibitive \cite{Duan-1992}.

To the scope a rule has been introduced to reject parameter-set where
each of them differs from initial and extreme values:
\begin{center}
\textbf{rule-b:} $|pi-p_{ini}|>\delta\cdot p_{ini}\quad OR\quad|pi-p_{min}|>\delta\cdot p_{ini}\quad OR\quad|pi-p_{max}|>\delta\cdot p_{ini}$
\par\end{center}

The set of parameters collected by data subsets have been finally
tested for normality by Jarque-Bera test, after cutting the edges
by different percentiles values. The considered window were $2-98%\ensuremath{\%}
_{ile}$, $5-95%\ensuremath{\%}
_{ile}$ and $10-90%\ensuremath{\%}
_{ile}$. Finally a comparison between parameter population have been performed
by a Principal Component Analysis (PCA) to assess if the soil of the
4 plots are really different.
\begin{description}
\item [{Comparison}] to experimental values of parameters have been only
performed for bulk density, which have been obtained in the field
with the classical core method, with samples taken horizontally from
a ditch (about 50cm diameter, 50cm depth) excavated inside each of
the 4 main plots, by steel cylinders of $100cm^{3}$ , with 2 replicates
for each depth.
\item [{Code}] has been developed in Matlab (R2011b, The Mathworks, Inc.).
For the NLF we used the embedded procedure \textbf{\textit{sqcurvefit}},\textbf{
}with the (default) 'trust-region-reflective' algorith\textbf{:} optimset('TolFun',1e-6,'TolX',1e-5,
'Algorithm','trust-region-reflective'). Jarque-Bera test and PCA are
also embedded in Matlab by the functions \textbf{\textit{jbtest}}
and\textit{ }\textbf{\textit{princomp}} functions respectively.
\end{description}

\section*{Results}

\begin{figure}[h]
\begin{centering}
\begin{tabular}{cc}
\includegraphics[width=8cm]{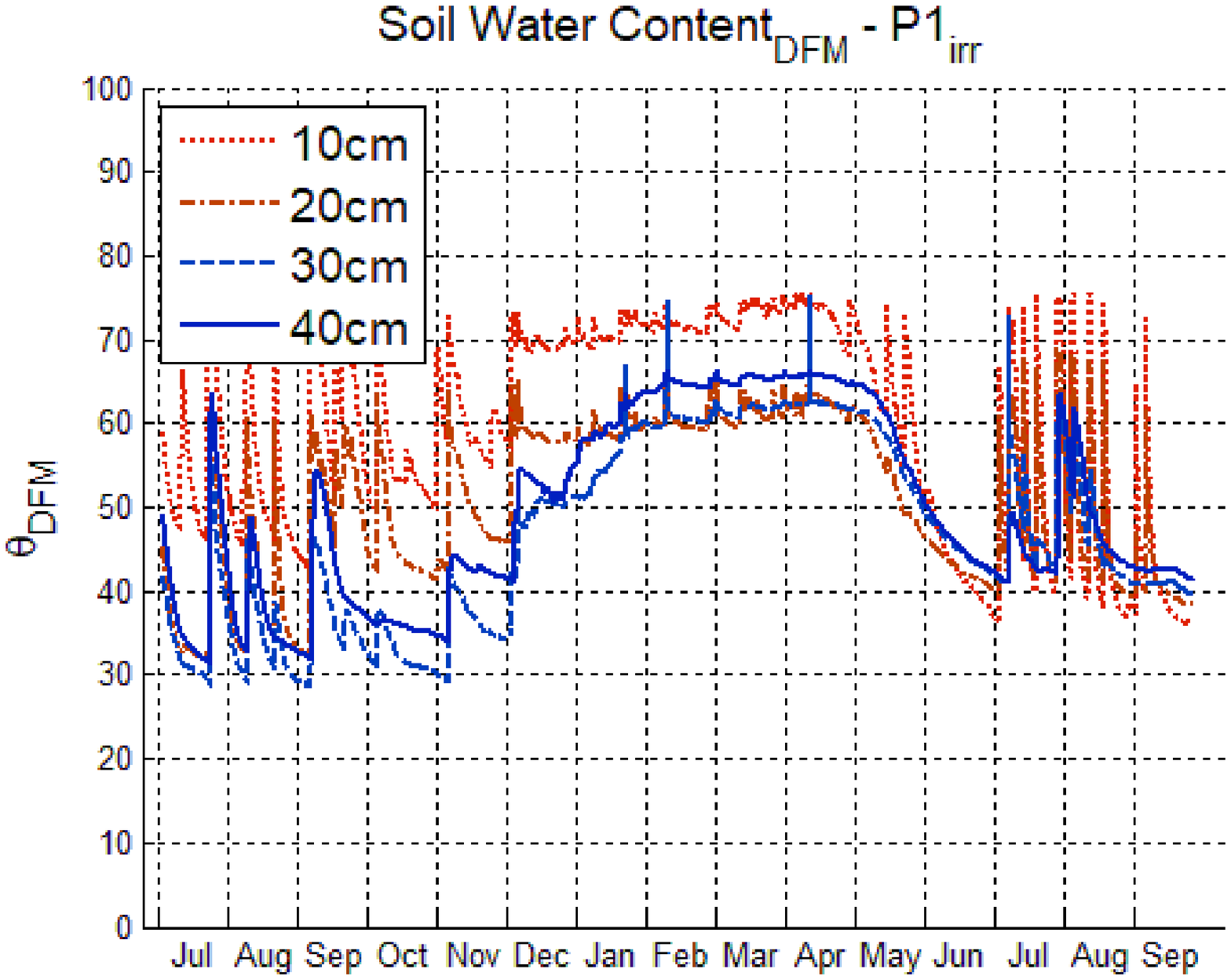} & \includegraphics[width=8cm]{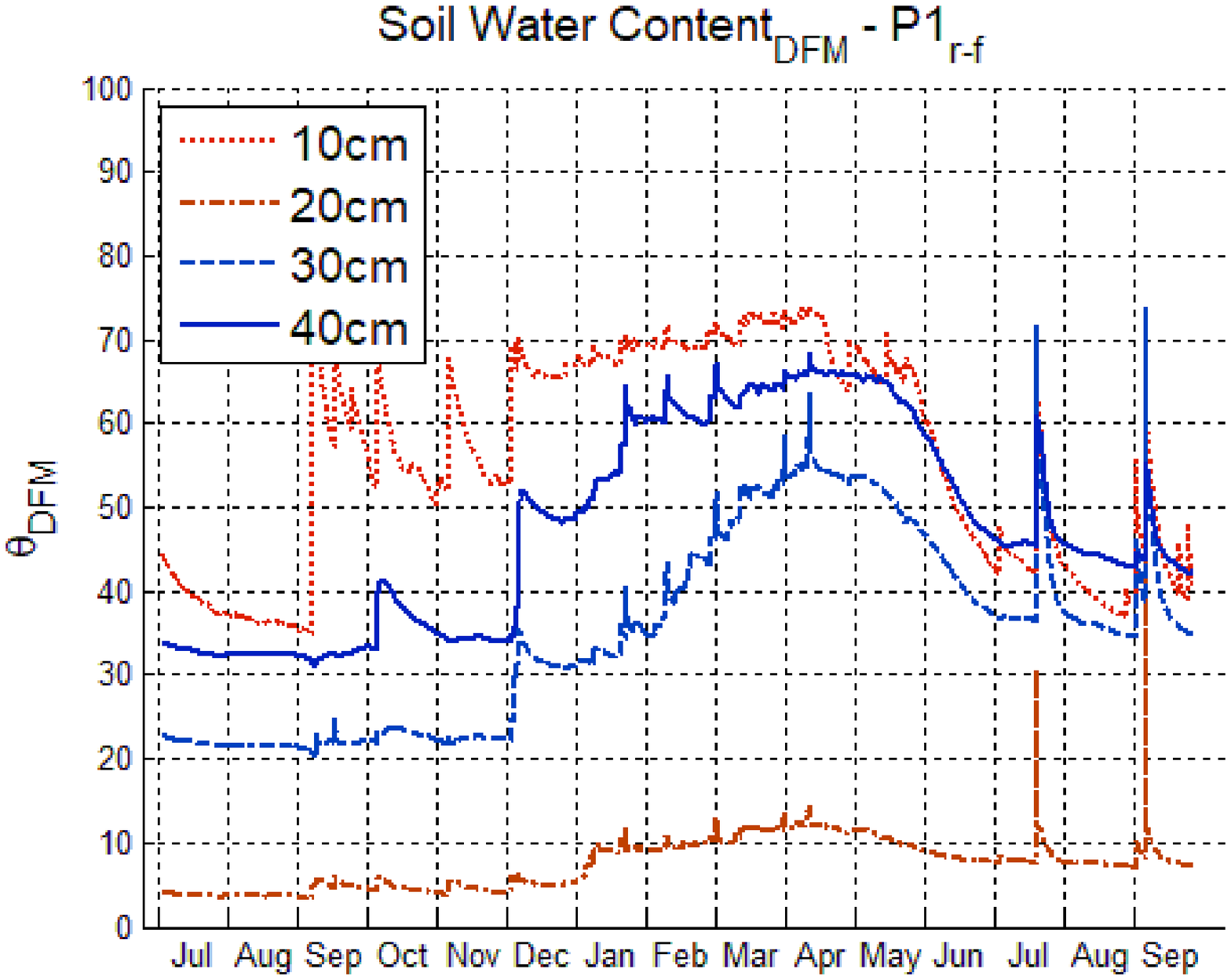}\tabularnewline
\includegraphics[width=8cm]{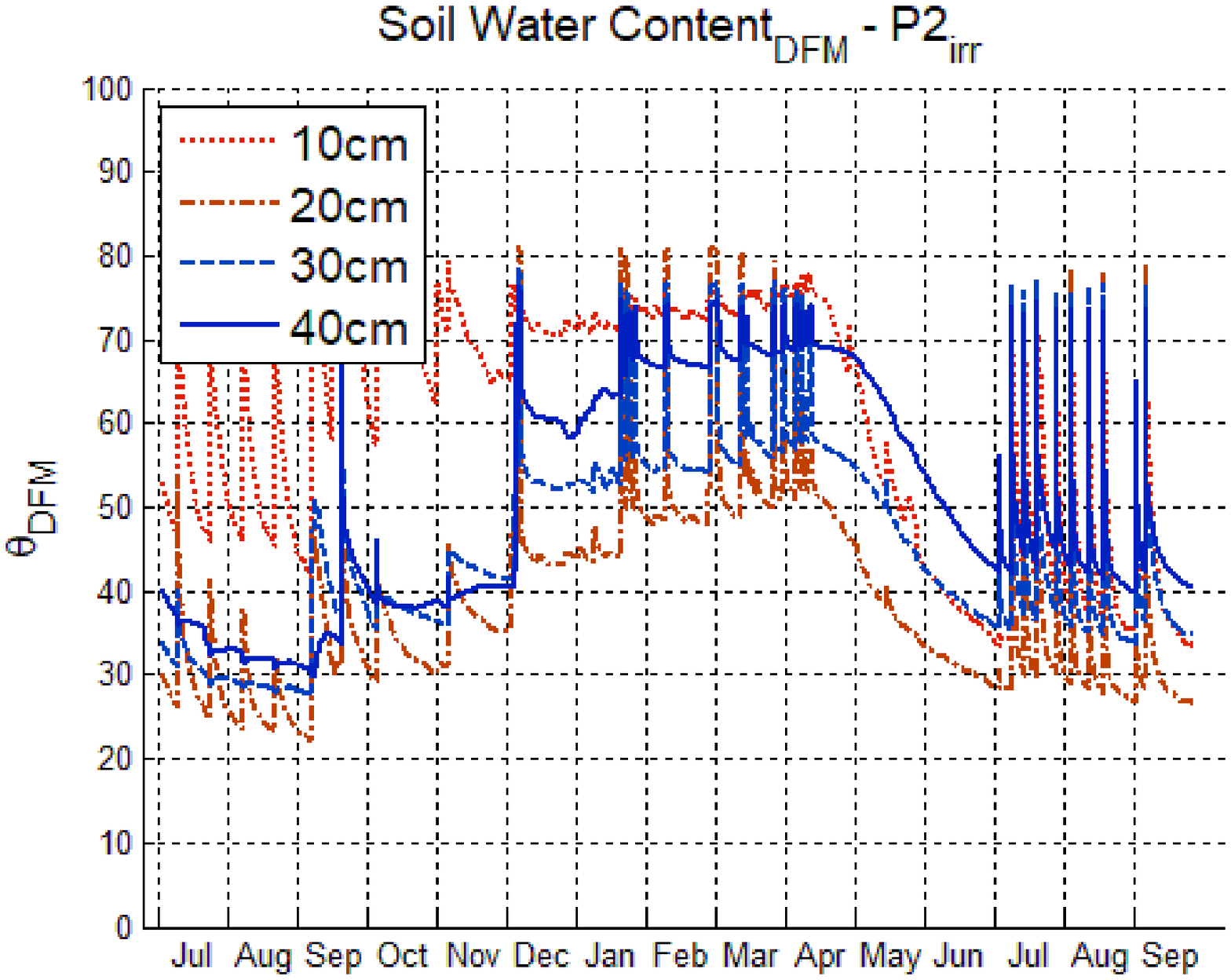} & \includegraphics[width=8cm]{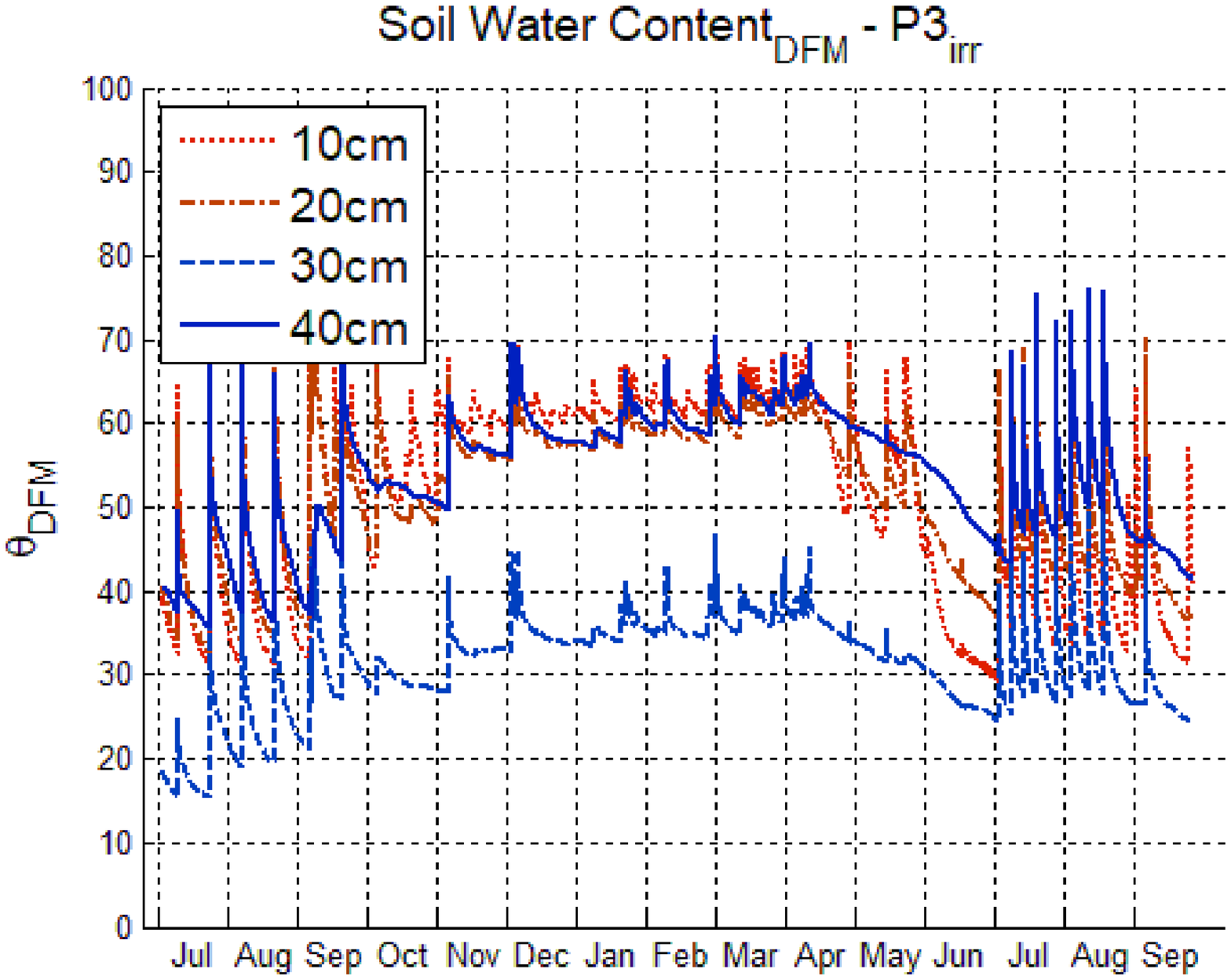}\tabularnewline
\includegraphics[width=8cm]{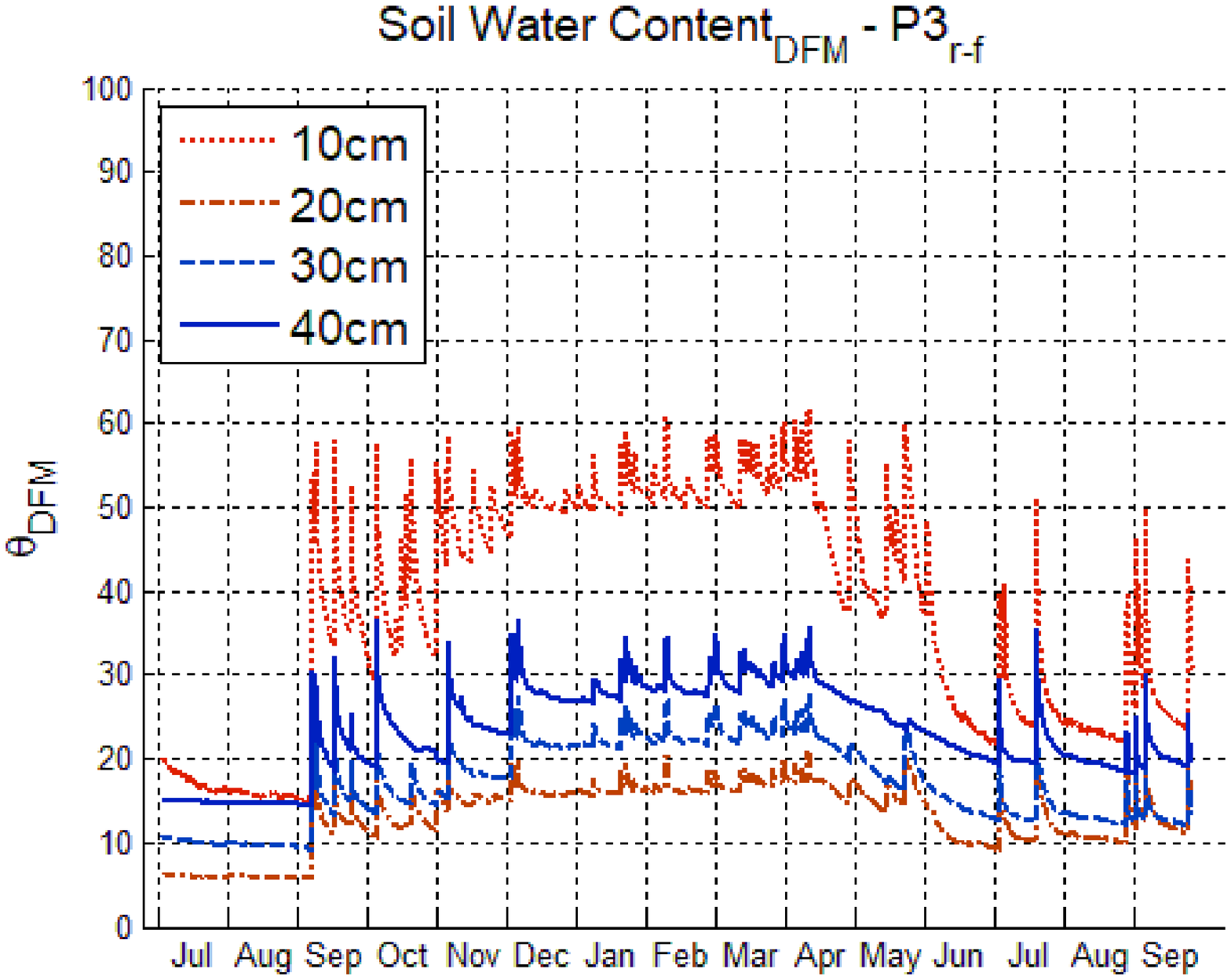} & \includegraphics[width=8cm]{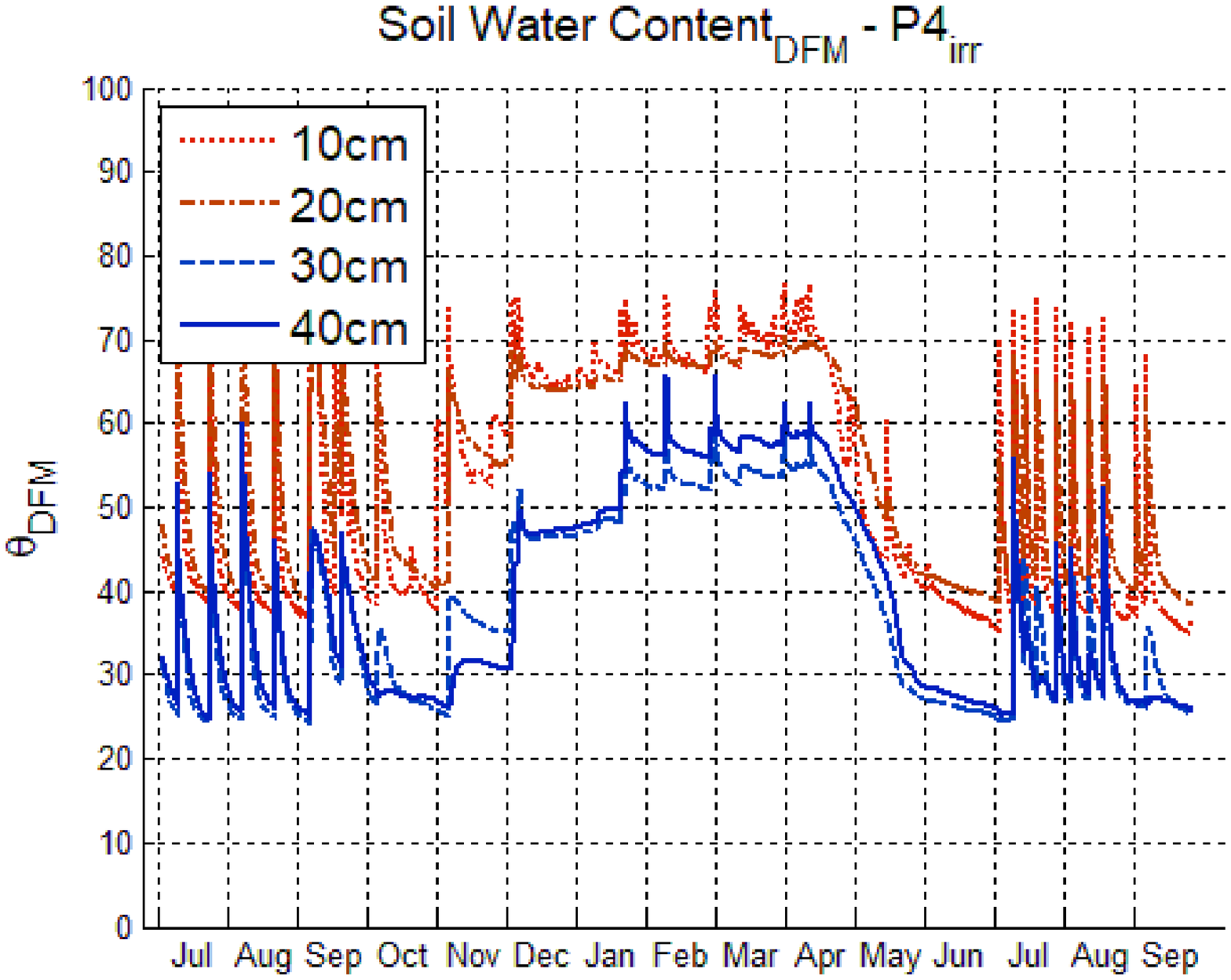}\tabularnewline
\end{tabular}
\par\end{centering}
\caption{\label{fig:2}DFM's SWC records for the 6 plots}
\end{figure}

Hourly records of SWC ( $\theta_{DFM}$) for the 6 plots (4 irrigated+2
r-f sub-plots) are shown in \ref{fig:2} where a marked seasonal trends
of soil moisture is visible at every depths.

Irrigated plots are easily recognizable from regular peaks in July
an August, which rapidly decrease because of redistribution characterizing
high conductive soils.

Inversion results are shown in tables \ref{tab:3},\ref{tab:4}, and\ref{tab:5}.
Table \ref{tab:3} reports the number of data-sets obtained for each
plot by \textbf{rule-a }used for inversion (\textbf{$n_{pre}$)},
the number of subsets passing \textbf{rule-b} and staying within the
3 percentile window considered, together with the relative J-B test
response\textbf{. }

\begin{table}
\begin{centering}
\begin{tabular}{ccccccccrcc}
\hline 
\textbf{plot} & \textbf{$n_{pre}$} &  & \textbf{$n_{pos}$} & \textbf{$J-B$} &  & \textbf{$n_{pos}$} & \textbf{$J-B$} &  & \textbf{$n_{pos}$} & \textbf{$J-B$}\tabularnewline
\hline 
 &  &  & \multicolumn{2}{c}{$2-98\%%_{ile}
$} &  & \multicolumn{2}{c}{$5-95%\%{}_{ile}
$} &  & \multicolumn{2}{r}{$10-90\%%_{ile}
$}\tabularnewline
\hline 
$1_{r-f}$ & \textbf{\footnotesize{}29} &  & {\footnotesize{}7} & {\footnotesize{}0} &  & {\footnotesize{}7} & {\footnotesize{}0} &  & \textbf{\footnotesize{}2} & \textbf{\footnotesize{}1}\tabularnewline
\hline 
$1_{irr}$ & \textbf{\footnotesize{}100} &  & \textbf{\footnotesize{}17} & \textbf{\footnotesize{}1} &  & \textbf{\footnotesize{}5} & \textbf{\footnotesize{}1} &  & \textbf{\footnotesize{}2} & \textbf{\footnotesize{}1}\tabularnewline
\hline 
$2_{irr}$ & \textbf{\footnotesize{}98} &  & {\footnotesize{}24} & {\footnotesize{}0} &  & \textbf{\footnotesize{}13} & \textbf{\footnotesize{}1} &  & \textbf{\footnotesize{}10} & \textbf{\footnotesize{}1}\tabularnewline
\hline 
$3_{r-f}$ & \textbf{\footnotesize{}77} &  & {\footnotesize{}32} & {\footnotesize{}0} &  & {\footnotesize{}24} & {\footnotesize{}0} &  & {\footnotesize{}7} & {\footnotesize{}0}\tabularnewline
\hline 
$3_{irr}$ & \textbf{\footnotesize{}141} &  & {\footnotesize{}38} & {\footnotesize{}0} &  & {\footnotesize{}30} & {\footnotesize{}0} &  & \textbf{\footnotesize{}13} & \textbf{\footnotesize{}1}\tabularnewline
\hline 
$4_{irr}$ & \textbf{\footnotesize{}127} &  & {\footnotesize{}42} & {\footnotesize{}0} &  & {\footnotesize{}31} & {\footnotesize{}0} &  & \textbf{\footnotesize{}19} & \textbf{\footnotesize{}1}\tabularnewline
\hline 
\end{tabular}
\par\end{centering}
\caption{\label{tab:3}Data sets available from data after filtering of rule-a
and rule-b with different percentile windows}
\end{table}

It can be seen that, as window becomes more selective, J-B test also
results positive (normal distribution) but population reduces sensibly,
therefore for the description of values (tables \ref{tab:4},\ref{tab:5})
the compromise window ($5-95%\%_{ile}
$) has been chosen. Table \ref{tab:4} shows the average values of
parameters. Columns 1 to 4 refer to bulk densities, with a value ranging
from $1.08$ to $1.32$: largest values are reached in subsurface
layers ($10-20cm$ and $20-30cm$).

About the parameters relative to WRC and UCC, despite of the result
of J-B test, their values are quite close to one another, strengthening
the idea that the plots have the same soil, and that any difference
should be ascribed to spatial variability. 

\begin{table}
\begin{centering}
\begin{tabular}{crrrrrrrrrrr}
\hline 
\textbf{plot} & \textbf{$\rho_{aps-10}$} & \textbf{$\rho_{aps-20}$} & \textbf{$\rho_{aps-30}$} & \textbf{$\rho_{aps-40}$} & \textbf{$\theta_{R}$} & \textbf{$a_{d}$} & \textbf{$a_{w}$} & \textbf{$n$} & \textbf{$K_{\mu}$} & \textbf{$c$} & \textbf{$b$}\tabularnewline
\hline 
 & {\footnotesize{}$(kg/m^{3})$} & {\footnotesize{}$(kg/m^{3})$} & {\footnotesize{}$(kg/m^{3})$} & {\footnotesize{}$(kg/m^{3})$} & {\footnotesize{}$(m^{3}/m^{3})$} & {\footnotesize{}$(-)$} & {\footnotesize{}$(-)$} & {\footnotesize{}$(-)$} & {\footnotesize{}$(cm/h)$} & {\footnotesize{}$(-)$} & {\footnotesize{}$(-)$}\tabularnewline
\hline 
$1_{r-f}$ & {\footnotesize{}1.09} & {\footnotesize{}1.32} & {\footnotesize{}1.06} & {\footnotesize{}1.01} & {\footnotesize{}0.019} & {\footnotesize{}0.033} & {\footnotesize{}0.091} & {\footnotesize{}1.65} & {\footnotesize{}0.51} & {\footnotesize{}0.67} & {\footnotesize{}26.8}\tabularnewline
\hline 
$1_{irr}$ & {\footnotesize{}1.10} & {\footnotesize{}1.13} & {\footnotesize{}1.13} & {\footnotesize{}1.10} & {\footnotesize{}0.027} & {\footnotesize{}0.028} & {\footnotesize{}0.053} & {\footnotesize{}1.78} & {\footnotesize{}0.50} & {\footnotesize{}0.60} & {\footnotesize{}23.1}\tabularnewline
\hline 
$2_{irr}$ & {\footnotesize{}1.10} & {\footnotesize{}1.25} & {\footnotesize{}1.15} & {\footnotesize{}1.07} & {\footnotesize{}0.027} & {\footnotesize{}0.031} & {\footnotesize{}0.068} & {\footnotesize{}1.50} & {\footnotesize{}0.57} & {\footnotesize{}0.69} & {\footnotesize{}25.2}\tabularnewline
\hline 
$3_{r-f}$ & {\footnotesize{}1.08} & {\footnotesize{}1.21} & {\footnotesize{}1.10} & {\footnotesize{}1.16} & {\footnotesize{}0.022} & {\footnotesize{}0.030} & {\footnotesize{}0.069} & {\footnotesize{}1.96} & {\footnotesize{}0.65} & {\footnotesize{}0.76} & {\footnotesize{}23.1}\tabularnewline
\hline 
$3_{irr}$ & {\footnotesize{}1.19} & {\footnotesize{}1.13} & {\footnotesize{}1.30} & {\footnotesize{}1.07} & {\footnotesize{}0.041} & {\footnotesize{}0.031} & {\footnotesize{}0.077} & {\footnotesize{}1.54} & {\footnotesize{}0.61} & {\footnotesize{}0.73} & {\footnotesize{}25.0}\tabularnewline
\hline 
$4_{irr}$ & {\footnotesize{}1.14} & {\footnotesize{}1.08} & {\footnotesize{}1.26} & {\footnotesize{}1.25} & {\footnotesize{}0.031} & {\footnotesize{}0.032} & {\footnotesize{}0.071} & {\footnotesize{}1.56} & {\footnotesize{}0.62} & {\footnotesize{}0.64} & {\footnotesize{}24.6}\tabularnewline
\hline 
\end{tabular}
\par\end{centering}
\caption{\label{tab:4}Parameter averaged values obtained by the inversion
technique}
\end{table}

Standard deviations are reported in \ref{tab:5}. 

\begin{table}[h]
\begin{centering}
\begin{tabular}{crrrrrrrrrrr}
\hline 
\textbf{plot} & \textbf{$\rho_{aps-10}$} & \textbf{$\rho_{aps-20}$} & \textbf{$\rho_{aps-30}$} & \textbf{$\rho_{aps-40}$} & \textbf{$\theta_{R}$} & \textbf{$a_{d}$} & \textbf{$a_{w}$} & \textbf{$n$} & \textbf{$K_{\mu}$} & \textbf{$c$} & \textbf{$b$}\tabularnewline
\hline 
 & {\footnotesize{}$(kg/m^{3})$} & {\footnotesize{}$(kg/m^{3})$} & {\footnotesize{}$(kg/m^{3})$} & {\footnotesize{}$(kg/m^{3})$} & {\footnotesize{}$(m^{3}/m^{3})$} & {\footnotesize{}$(-)$} & {\footnotesize{}$(-)$} & {\footnotesize{}$(-)$} & {\footnotesize{}$(cm/h)$} & {\footnotesize{}$(-)$} & {\footnotesize{}$(-)$}\tabularnewline
\hline 
$1_{r-f}$ & {\footnotesize{}0.13} & {\footnotesize{}0.06} & {\footnotesize{}0.03} & {\footnotesize{}0.06} & {\footnotesize{}0.013} & {\footnotesize{}0.005} & {\footnotesize{}0.03} & {\footnotesize{}0.38} & {\footnotesize{}0.23} & {\footnotesize{}0.13} & {\footnotesize{}3.2}\tabularnewline
\hline 
$1_{irr}$ & {\footnotesize{}0.05} & {\footnotesize{}0.07} & {\footnotesize{}0.04} & {\footnotesize{}0.07} & {\footnotesize{}0.017} & {\footnotesize{}0.005} & {\footnotesize{}0.03} & {\footnotesize{}0.22} & {\footnotesize{}0.22} & {\footnotesize{}0.19} & {\footnotesize{}5.0}\tabularnewline
\hline 
$2_{irr}$ & {\footnotesize{}0.04} & {\footnotesize{}0.01} & {\footnotesize{}0.02} & {\footnotesize{}0.04} & {\footnotesize{}0.016} & {\footnotesize{}0.003} & {\footnotesize{}0.01} & {\footnotesize{}0.20} & {\footnotesize{}0.13} & {\footnotesize{}0.09} & {\footnotesize{}2.9}\tabularnewline
\hline 
$3_{r-f}$ & {\footnotesize{}0.03} & {\footnotesize{}0.04} & {\footnotesize{}0.06} & {\footnotesize{}0.03} & {\footnotesize{}0.009} & {\footnotesize{}0.006} & {\footnotesize{}0.03} & {\footnotesize{}0.47} & {\footnotesize{}0.09} & {\footnotesize{}0.06} & {\footnotesize{}3.7}\tabularnewline
\hline 
$3_{irr}$ & {\footnotesize{}0.06} & {\footnotesize{}0.03} & {\footnotesize{}0.02} & {\footnotesize{}0.02} & {\footnotesize{}0.017} & {\footnotesize{}0.001} & {\footnotesize{}0.01} & {\footnotesize{}0.07} & {\footnotesize{}0.04} & {\footnotesize{}0.03} & {\footnotesize{}1.5}\tabularnewline
\hline 
$4_{irr}$ & {\footnotesize{}0.03} & {\footnotesize{}0.02} & {\footnotesize{}0.01} & {\footnotesize{}0.02} & {\footnotesize{}0.011} & {\footnotesize{}0.003} & {\footnotesize{}0.02} & {\footnotesize{}0.08} & {\footnotesize{}0.07} & {\footnotesize{}0.10} & {\footnotesize{}3.1}\tabularnewline
\hline 
\end{tabular}
\par\end{centering}
\caption{\label{tab:5}Parameter standard deviation values obtained by the
inversion technique}
\end{table}

To corroborate the hypothesis of a common soil, a PCA has been performed
on the last 7 parameters (bulk density has been excluded as it depends
on depth also), using the 110 parameter sets. From PCA it resulted
that the first principal component is responsible of $48.3%\ensuremath{\%}
$ of variability with an eigenvector (table \ref{tab:6}) ascribing
such variability to the main shape factors of WRC ($a_{d}$ and $n$).
Component \#2 and \#3 are responsible of respectively $16.8%\ensuremath{\%}
$ and $15.7%\ensuremath{\%}
$; the other 4 components are of minor importance.

Compaction factor $c$, influenceing conductivity, is relevant in
component \#3, \#5 and \#6, whereas \textbf{$a_{w}/a_{d}$ }related
to hysteresis, mainly affects \#4, to which is ascribed $8.6\%$ of
global variance.

\begin{table}
\begin{centering}
\begin{tabular}{cclllllll}
\textbf{component} & \textbf{var} & \textbf{$\theta_{R}$} & \textbf{$a_{d}$} & \textbf{$a_{w}/a_{d}$} & \textbf{$n$} & \textbf{$K_{\mu}$} & \textbf{$c$} & \textbf{$b$}\tabularnewline
1 & \textbf{\footnotesize{}48.3} & {\footnotesize{}-0.019} & \textbf{\footnotesize{}0.829} & {\footnotesize{}-0.162} & \textbf{\footnotesize{}0.511} & {\footnotesize{}-0.067} & {\footnotesize{}-0.050} & {\footnotesize{}-0.134}\tabularnewline
2 & \textbf{\footnotesize{}16.8} & \textbf{\footnotesize{}0.431} & {\footnotesize{}0.235} & {\footnotesize{}0.187} & {\footnotesize{}-0.250} & \textbf{\footnotesize{}0.779} & {\footnotesize{}0.086} & {\footnotesize{}-0.215}\tabularnewline
3 & \textbf{\footnotesize{}15.7} & \textbf{\footnotesize{}0.425} & {\footnotesize{}0.239} & {\footnotesize{}0.202} & \textbf{\footnotesize{}-0.469} & \textbf{\footnotesize{}-0.5091} & \textbf{\footnotesize{}-0.452} & {\footnotesize{}-0.193}\tabularnewline
4 & \textbf{\footnotesize{}8.6} & {\footnotesize{}0.083} & {\footnotesize{}-0.124} & \textbf{\footnotesize{}0.863} & \textbf{\footnotesize{}0.475} & {\footnotesize{}-0.049} & {\footnotesize{}-0.054} & {\footnotesize{}0.033}\tabularnewline
5 & \textbf{\footnotesize{}5.2} & \textbf{\footnotesize{}-0.462} & {\footnotesize{}0.259} & {\footnotesize{}0.180} & {\footnotesize{}-0.247} & {\footnotesize{}0.264} & \textbf{\footnotesize{}-0.495} & \textbf{\footnotesize{}0.558}\tabularnewline
6 & \textbf{\footnotesize{}3.2} & \textbf{\footnotesize{}-0.402} & {\footnotesize{}0.315} & {\footnotesize{}0.343} & \textbf{\footnotesize{}-0.411} & {\footnotesize{}-0.190} & \textbf{\footnotesize{}0.633} & {\footnotesize{}-0.121}\tabularnewline
7 & \textbf{\footnotesize{}2.2} & \textbf{\footnotesize{}0.501} & {\footnotesize{}0.139} & {\footnotesize{}-0.040} & {\footnotesize{}-0.004} & {\footnotesize{}-0.144} & {\footnotesize{}0.369} & \textbf{\footnotesize{}0.756}\tabularnewline
\end{tabular}
\par\end{centering}
\caption{\label{tab:6}Results of the PCA analysis in terms of component effects
on variances and eigenvectors composition; larger values are put in
bold}
\end{table}

\begin{figure}[h]
\centering{}%
\begin{tabular}{cc}
\includegraphics[width=8cm]{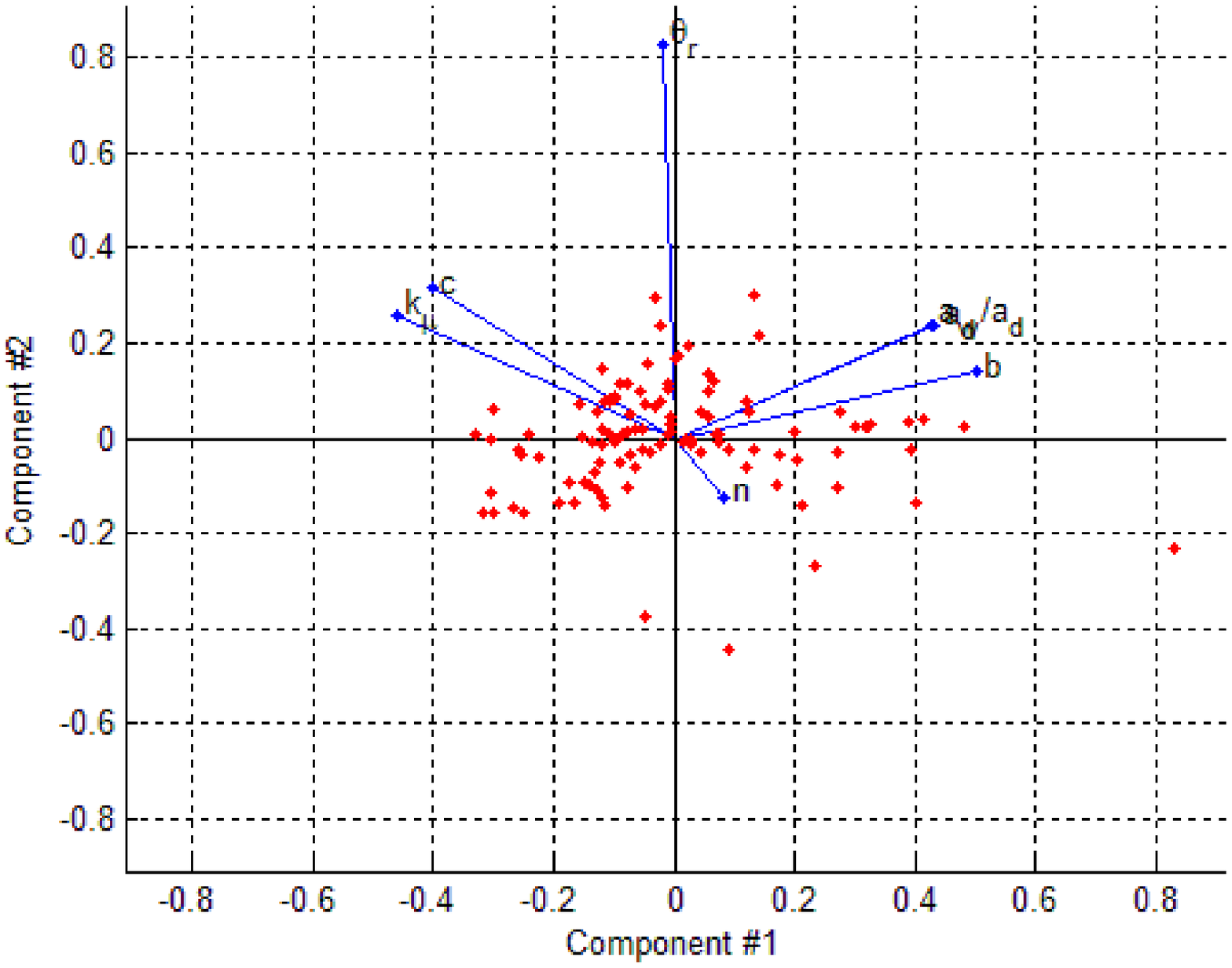} & \includegraphics[width=8cm]{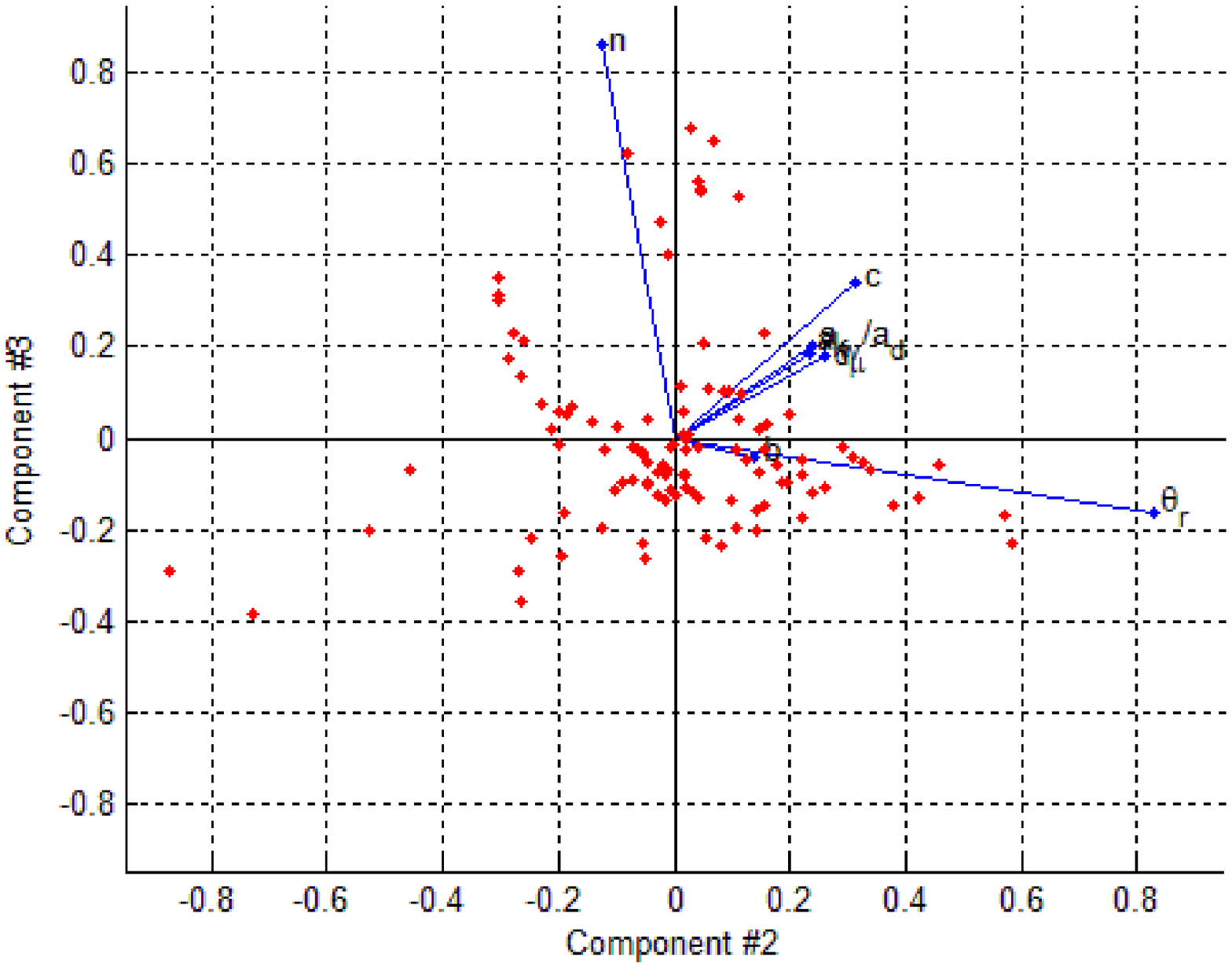}\tabularnewline
\includegraphics[width=8cm]{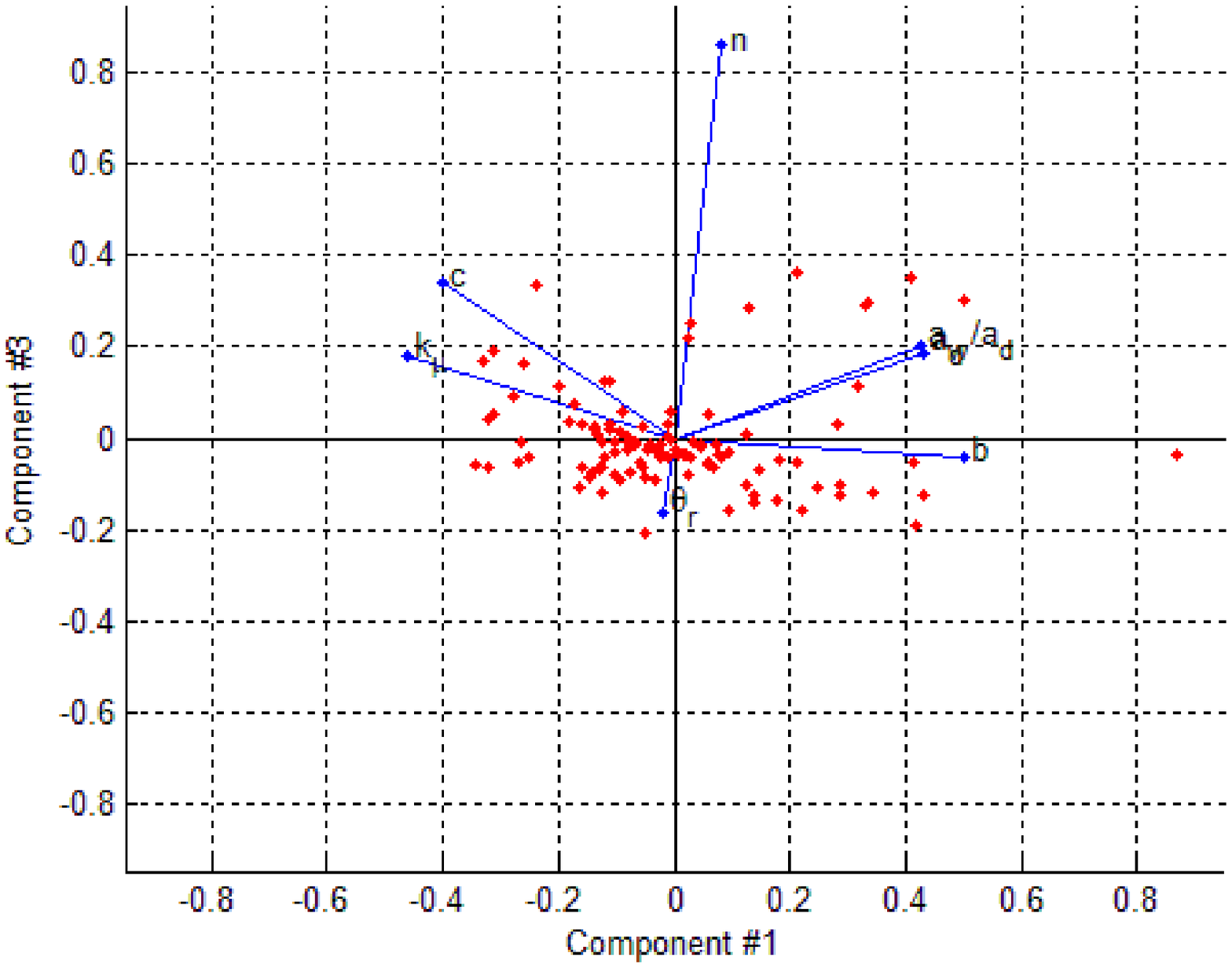} & \includegraphics[width=8cm]{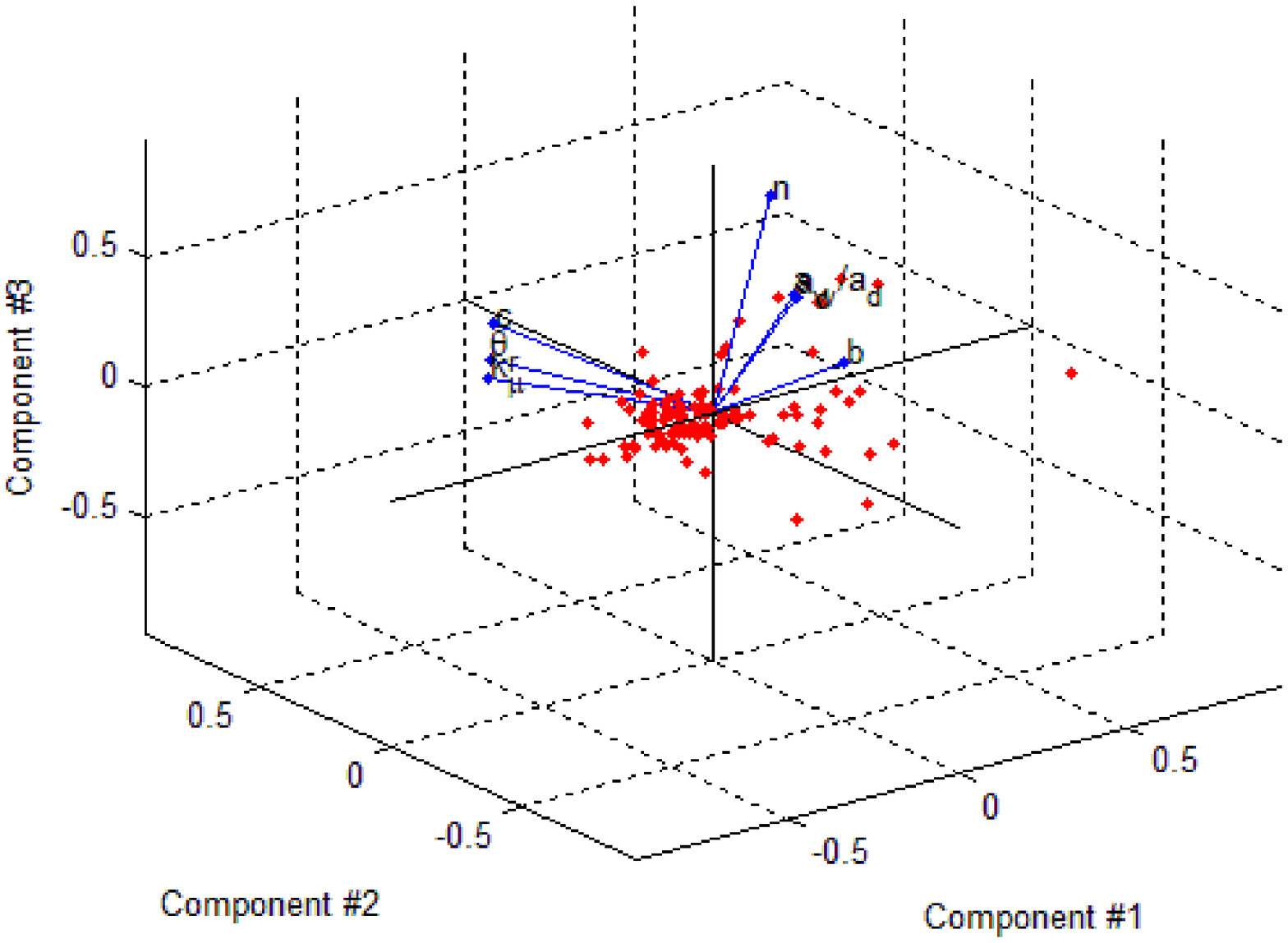}\tabularnewline
\end{tabular}\caption{\label{Fig:3}Plot of the cases (red points) and of main 2 components}
\end{figure}

A plot of the three main components is drawn in figure \ref{Fig:3}
show there is no evidence of clustering.

\begin{figure}[H]
\centering{}%
\begin{tabular}{c}
\includegraphics[width=12cm]{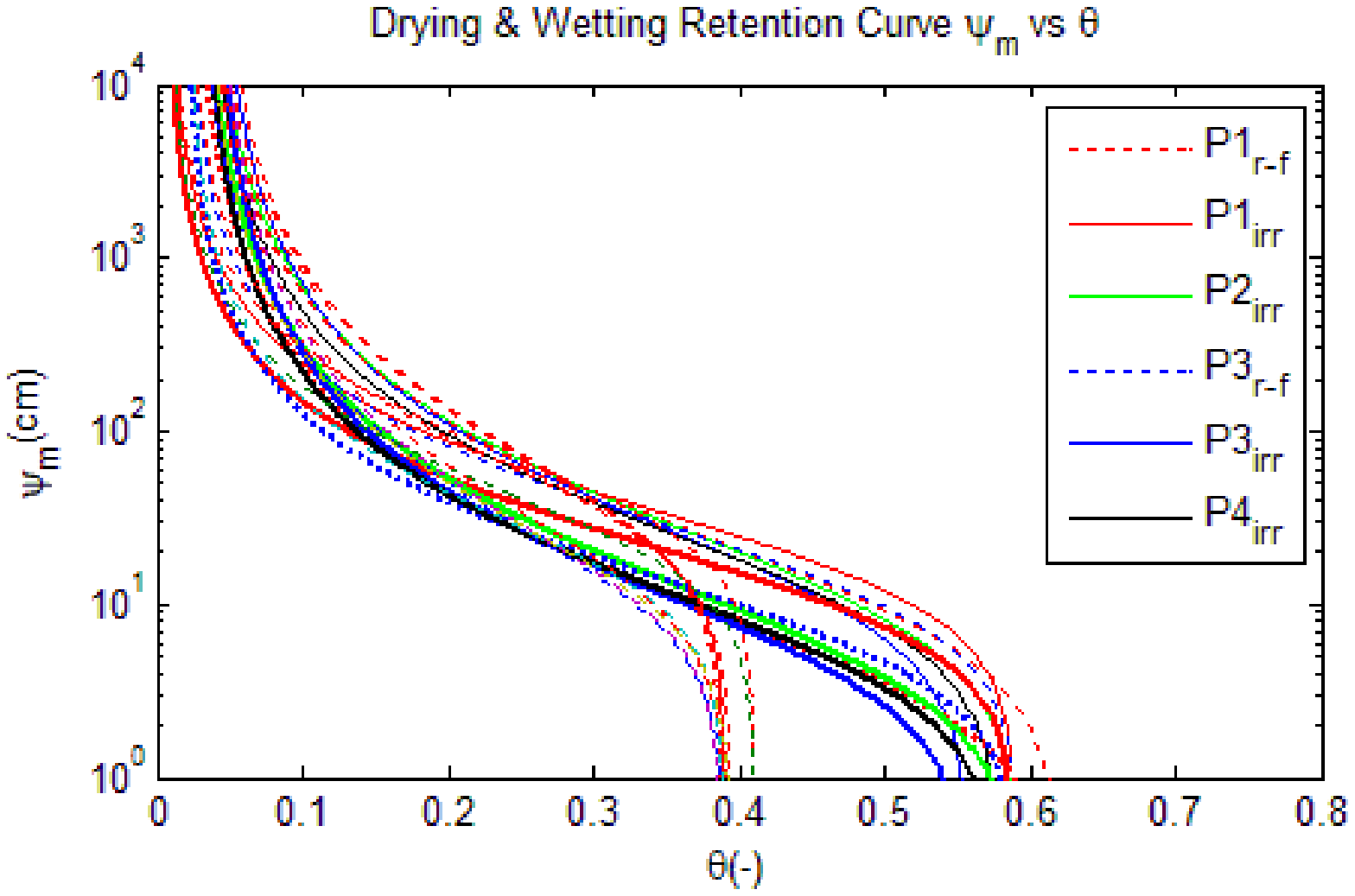}\tabularnewline
\includegraphics[width=12cm]{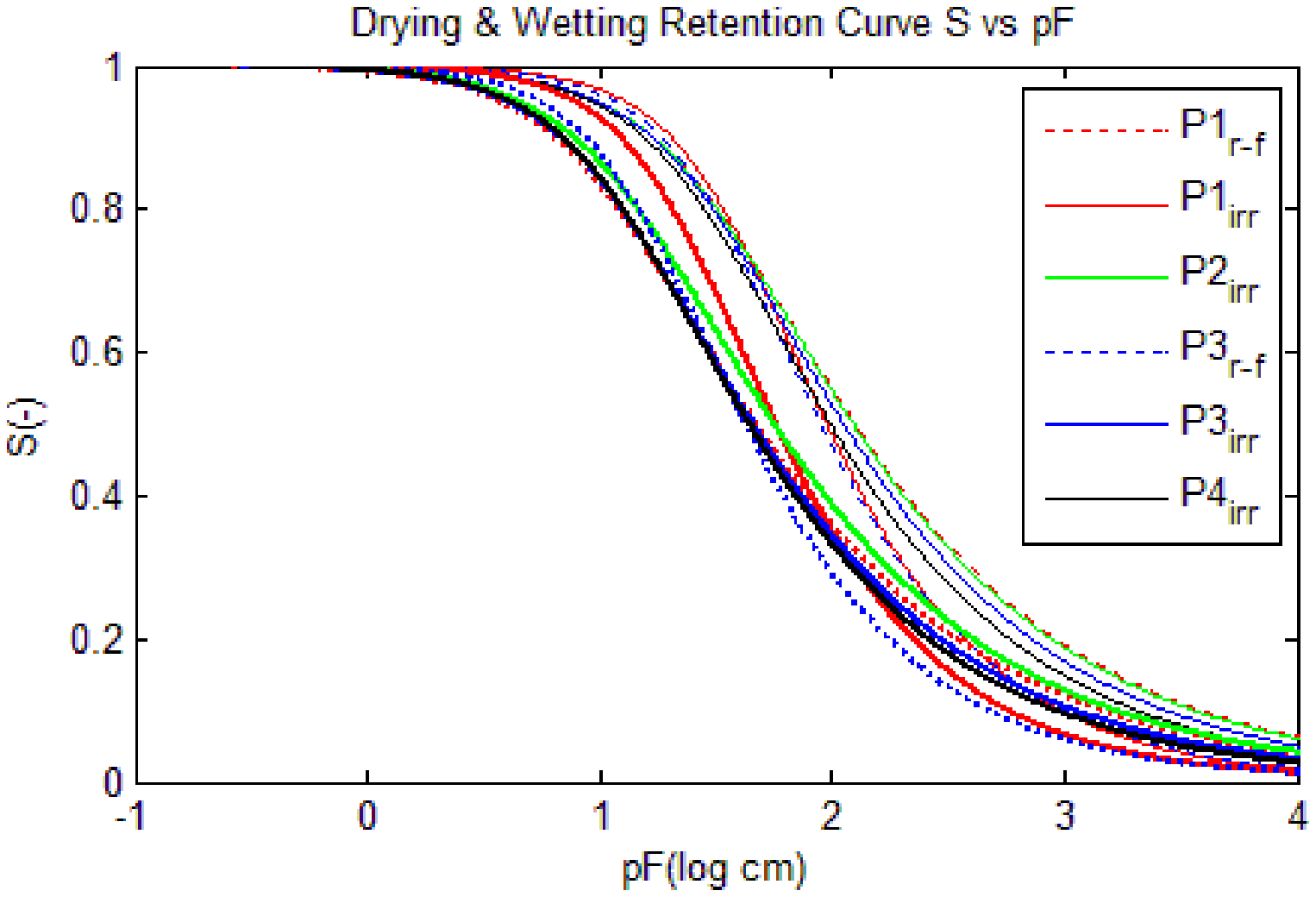}\tabularnewline
\end{tabular}\caption{\label{fig:4}WRCs for the 6 plots: continuous lines are for irrigated
plots, dashed lines for rain-fed plots. Bold lines are wetting curves.In
the top graph, both un-stretched and stretched (for top soil layer)
WRCs are shown.}
\end{figure}

Figure \ref{fig:4} shows the WRCs for the 6 cases.

\begin{figure}[H]
\centering{}\includegraphics[width=14cm]{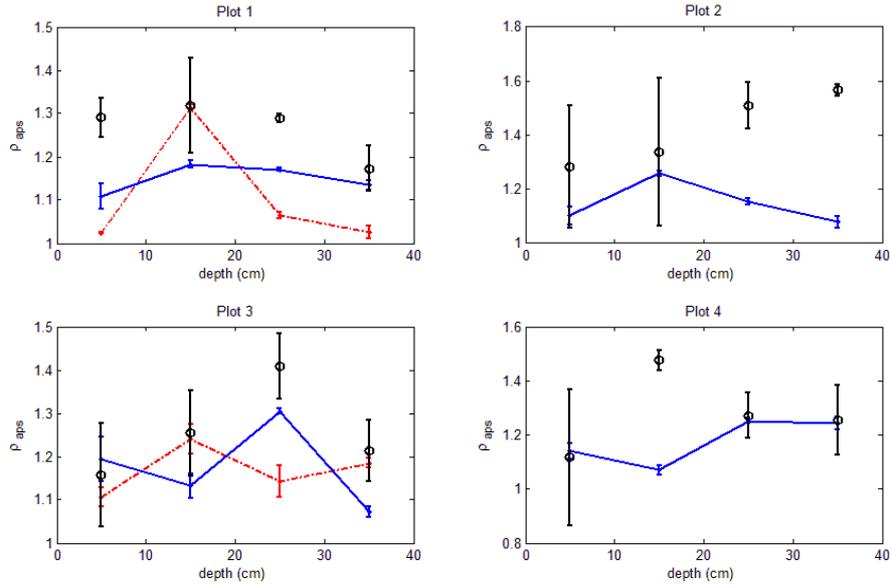}\caption{\label{fig:5}Estimated and measured bulk densities (BD) for the 4
plots; plots 1,3 have rain-fed (red lines) and irrigated conditions,
plots 2,4 only irrigated (blue lines). Circles represent measured
data. Error bars stand for standard deviations. }
\end{figure}

Experimental evaluation of bulk densities show high error bars of
experimental values, due to operative difficulties intrinsic of the
method, hard to be overcome in natural soils, due to roots and skeleton.
Inverse problem solving, if from the one side seems to underestimate
BD values, it may also confirm a typical problem of core method, soil
compression due to dragging effect of the cylinder wall. Model also
emphasizes bulk density differences along soil depth, most of times
in agreement with those observed with core method, as represented
in Figure \ref{fig:5}.

Indirect validation of parameters have been performed using the SWP
as a driving variable in a recently developed model to simulate mycelium
growth in the same plots \cite{Iotti-2018}.

\section*{Conclusions}

Inversion technique developed in this study, obtained by integration
of a refined soil water dynamical model in a classical non-linear
fitting method, is used to derive in-situ soil hydrological parameters
from soil moisture multi-depth probe records. The features included,
namely soil hysteresis, dependence of conductivity to bulk-density,
and inclusion of calibration curve confirm that inversion procedures
can give parameter with a validity both comparable to experimental
methods, and adherent to a real-world soil system.

The analysis also confirms that convergence strongly depended on the
size of available data records: in fact in general data can not guarantee
to own the information required from problem inversion.

From PCA analysis emerges that the plots analyzed corresponds to the
same soil, and that spatial variability affects the value of each
of parameters used to represent the soil.

Finally, though the methodology gave prove of robustness, it is required
to be assessed jet on a set of sufficiently different soils jet. 

The method can be extended to a wide class of probes, including the
cheapest ones today wide-spreading under the push of Internet of Things
(IoT), which rapidly increase the number of highly connected low cost
specialized devices placed everywhere, and that will give the opportunity
to enrich enormously the knowledge about soils and its reliability.

\bibliographystyle{amsplain}
\bibliography{D:/BIB/SOIL-InverseProblem}

\providecommand{\bysame}{\leavevmode\hbox to3em{\hrulefill}\thinspace}
\providecommand{\MR}{\relax\ifhmode\unskip\space\fi MR }
\providecommand{\MRhref}[2]{%
  \href{http://www.ams.org/mathscinet-getitem?mr=#1}{#2}
}
\providecommand{\href}[2]{#2}
\begin{thebibliography}{10}

\bibitem{DFM-2010}
AAVV{\_}DFM, \emph{{The DFM continuous logging soil moisture probe - Functional
  specifications}}, Tech. report, 2010.

\bibitem{Conn-2000}
Andrew~R. Conn, Nicholas I.~M. Gould, and Philippe~L. Toint,
  \emph{{Trust-region methods}}, Society for Industrial and Applied
  Mathematics, 2000.

\bibitem{Duan-1992}
Qingyun Duan, Soroosh Sorooshian, and Vijai Gupta, \emph{{Effective and
  efficient global optimization for conceptual rainfall-runoff models}}, Water
  Resources Research \textbf{28} (1992), no.~4, 1015--1031.

\bibitem{Durner-2005}
Wolfgang Durner and Kai Lipsius, \emph{{Determining Soil Hydraulic
  Properties}}, Encyclopedia of Hydrological Sciences, John Wiley {\&} Sons,
  Ltd, Chichester, UK, oct 2005.

\bibitem{Elmaloglou-2008}
S.~Elmaloglou and E.~Diamantopoulos, \emph{{The effect of hysteresis on
  three-dimensional transient water flow during surface trickle irrigation}},
  Irrigation and Drainage \textbf{57} (2008), no.~1, 57--70.

\bibitem{Iotti-2018}
Mirco Iotti, Pamela Leonardi, Giuliano Vitali, and Alessandra Zambonelli,
  \emph{{Effect of summer soil moisture and temperature on the vertical
  distribution of Tuber magnatum mycelium in soil}}, submitted (2018).

\bibitem{Mjanyelwa-2016}
Nokhwezi Mjanyelwa, Zaid~A. Bello, Willnerie Greaves, and Leon~D. van Rensburg,
  \emph{{Precision and accuracy of DFM soil water capacitance probes to measure
  temperature}}, Computers and Electronics in Agriculture \textbf{125} (2016),
  no.~C, 125--128.

\bibitem{Mualem-1976}
Yechezkel Mualem, \emph{{A new model for predicting the hydraulic conductivity
  of unsaturated porous media}}, Water Resources Research \textbf{12} (1976),
  no.~3, 513--522.

\bibitem{Nelder-1965}
J.~A. Nelder and R.~Mead, \emph{{A Simplex Method for Function Minimization}},
  The Computer Journal \textbf{7} (1965), no.~4, 308--313.

\bibitem{Osunbitan-2005}
J.A. Osunbitan, D.J. Oyedele, and K.O. Adekalu, \emph{{Tillage effects on bulk
  density, hydraulic conductivity and strength of a loamy sand soil in
  southwestern Nigeria}}, Soil and Tillage Research \textbf{82} (2005), no.~1,
  57--64.

\bibitem{Shaykewitch-1970}
C.F. Shaykewitch, \emph{{Hydraulic Properties of Disturbed and Undisturbed
  Soils}}, Can.J.Soil Sci. \textbf{50} (1970), 430--437.

\bibitem{Tarantola-2005}
Albert Tarantola, \emph{{Inverse Problem Theory and Methods for Model Parameter
  Estimatio}}, SIAM-Society for Industrial and Applied Mathematics,
  Phyladelphia, 2005.

\bibitem{VanGenuchten-1980}
M.~Th. van Genuchten, \emph{{A Closed-form Equation for Predicting the
  Hydraulic Conductivity of Unsaturated Soils1}}, Soil Science Society of
  America Journal \textbf{44} (1980), no.~5, 892.

\bibitem{Vrugt-2008}
Jasper~A. Vrugt, Philip~H. Stauffer, Th. Wohling, Bruce~A. Robinson, and
  Velimir~V. Vesselinov, \emph{{Inverse Modeling of Subsurface Flow and
  Transport Properties: A Review with New Developments}}, Vadose Zone Journal
  \textbf{7} (2008), no.~2, 843.

\bibitem{Wijaya-2016}
M.~Wijaya and E.C. Leong, \emph{{Equation for unimodal and bimodal soil–water
  characteristic curves}}, Soils and Foundations \textbf{56} (2016), no.~2,
  291--300.

\end{thebibliography}

\subsection*{Credits}

The manuscript has been edited by Lyx ver.2.1.1, using as Document
Class the Article (Standard Class), references have been managed by
Mendeley ver.1.17.13 and JabRef 3.8.2. 

\subsection*{Appendix A - Calibration}

DFM probe calibration DFM probes estimate soil water content on the
basis of dielectric properties of wet soil. As they are mostly used
for a comparative use, within a water scheduling framework, they come
with a general purpose calibration which does not ensure a precise
evaluation of water content for local soil. Nevertheless the several
sensors DFM probes are equipped with (DFM400 has $4$ $10cm$-spaced
sensors) can be assumed to have an identical behavior, therefore the
calibration has been performed on one of them. To calibrate a sensor
the DFM probe have been placed in a plastic cylinder ($diam=12cm$,
$len=10cm$) so as to have a single sensor surrounded by soil with
a known density and water content (added to the soil after it be oven-dried).
The cylinder has a bottom with a central hole for the probe drilled
to drain excess water, whereas is open above and hanged to a scale,
allowing a gravitational determination of water content. The cylinder
is formerly filled with dry soil, then saturated and left drying in
a warm room. The trial has been repeated with different soils and
different density, allowing to obtain the following calibration expression:
\begin{equation}
\theta=[0.01\:\theta_{DFM}\:(1+0.0018\:(20-Ts))\lyxmathsym{\textendash}a_{1}\lyxmathsym{\textendash}a_{2}\rho_{aps}^{m}]/[a_{3}+a_{4}\rho_{aps}^{m}]^{1/m}
\end{equation}

where $\theta_{DFM}$ is the read value, Ts is soil temperature and
$\rho_{aps}$ the bulk density. Parameters, estimated by a trial and
error procedure, are: $a_{1}=46.2$,$a_{2}=46.7$, $a_{3}=-51.8$
, $a_{4}=53.1$, $m=0.2=1/5$ ,with $R^{2}=0.993$.\newpage{}

\subsection*{Appendix B - Numerical scheme}

The equations described above have been discretized as follows below
to get the estimated SWC $\bar{\theta}$ in the intermediate nodes
($j=2,3$; $20,30cm$):

\begin{eqnarray*}
\overline{\theta}_{i,j} & = & \frac{[K_{j-1}+K_{j}]}{2}\cdot\frac{[\psi_{i,j-1}-\psi_{i,j}-dz]}{dz}\cdot dt-\frac{[K_{j}+K_{j+1}]}{2}\cdot\frac{[\psi_{i,j}-\psi_{i,j+1}-dz]}{dz}\cdot dt\\
K_{j} & = & Ks_{j}\cdot S_{i,j}^{b}(1-(1-S_{i,j}^{1/m})^{m})^{2}\\
Ks_{j} & = & K_{\mu}\cdot(1+c\cdot[\rho_{aps-j}-1])\\
\psi_{i,j} & = & (S_{i,j}{}^{n/(1-n)}-1)^{(1/n)}/a
\end{eqnarray*}

To include hysteresis procedure, the parameter $a$ is assigned one
of two different values, depending on sign of water content change:

\begin{eqnarray*}
a= & \begin{cases}
a_{w}\Leftarrow\theta_{DFM;i}<\theta_{DFM;i+1}\\
a_{d}\Leftarrow\theta_{DFM;i}>\theta_{DFM;i+1}
\end{cases}
\end{eqnarray*}

and saturation $S$ is computed by the stretching function for the
given SWC:

\begin{eqnarray*}
S_{i,j} & = & erf((\theta_{i,j}-\theta_{R})/slp)/A\\
\theta_{i,j} & = & f[\theta_{DFM;i,j},T_{S-i,j},\rho_{aps-j}]\\
A & = & erf[(\theta_{S-j}-\theta_{R})/slp]\\
slp & = & (\theta_{e-j}-\theta_{R})/S_{e}\\
\theta_{e-j} & = & (\theta_{\mu}-\theta_{R})\cdot S_{e}+\theta_{R}\\
\theta_{\mu} & = & (1-\rho_{\mu}/\rho_{s})\\
\theta_{S-j} & = & (1-\rho_{aps-j}/\rho_{s})\\
\rho_{aps-j} & = & \rho_{\mu}-\delta\rho_{j}\\
S_{e} & = & 2^{(1-n)/n}
\end{eqnarray*}

where $\delta\rho_{1..4}$ , $\theta_{R}$, $a_{d}$ ,$a_{w}$, $n$,
$K_{\mu}$, $c$ ,$b$, are the fitted parameters.\newpage{}

\subsection*{Appendix C - Parameter Distributions}

In the following figures, are the values of parameter obtained from
model inversion for the 6 plots.

\begin{figure}[h]
\begin{centering}
\includegraphics[width=14cm]{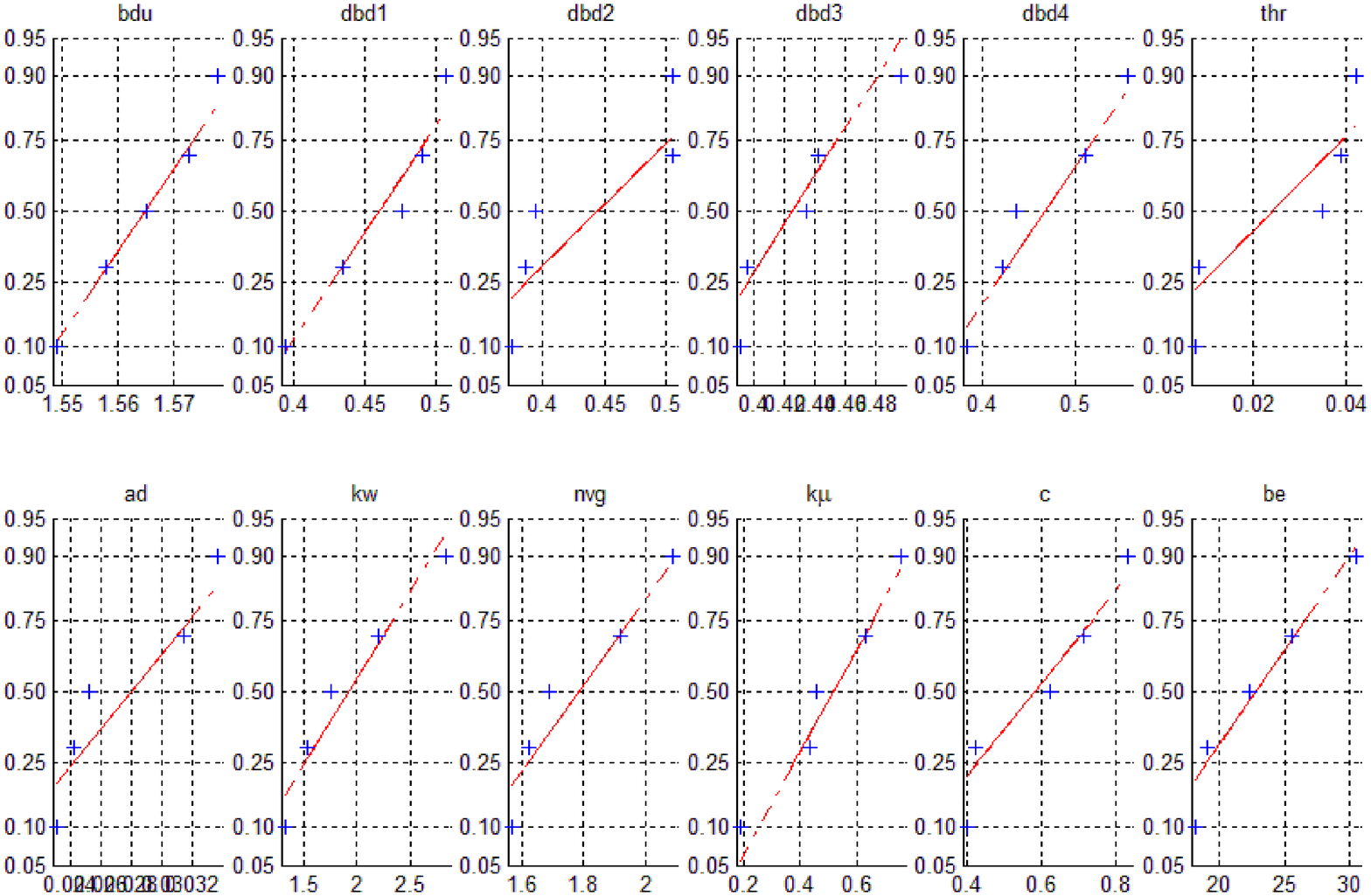}\caption{Parameter values for Plot 1 - irrigated}
\par\end{centering}
\end{figure}

\begin{figure}[h]
\centering{}\includegraphics[width=14cm]{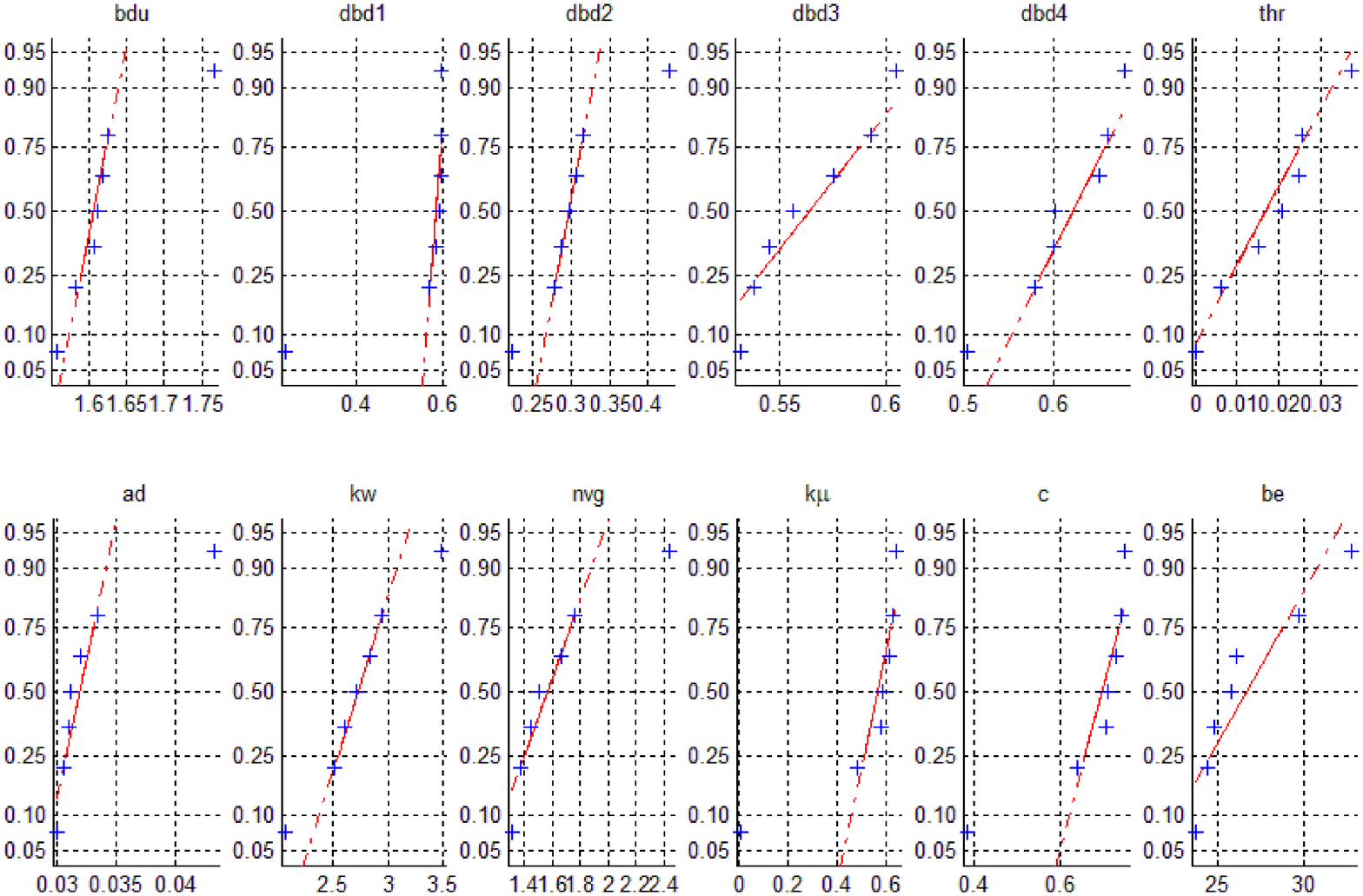}\caption{Parameter values for Plot 1 - rainf-fed}
\end{figure}

\begin{figure}[h]
\centering{}\includegraphics[width=14cm]{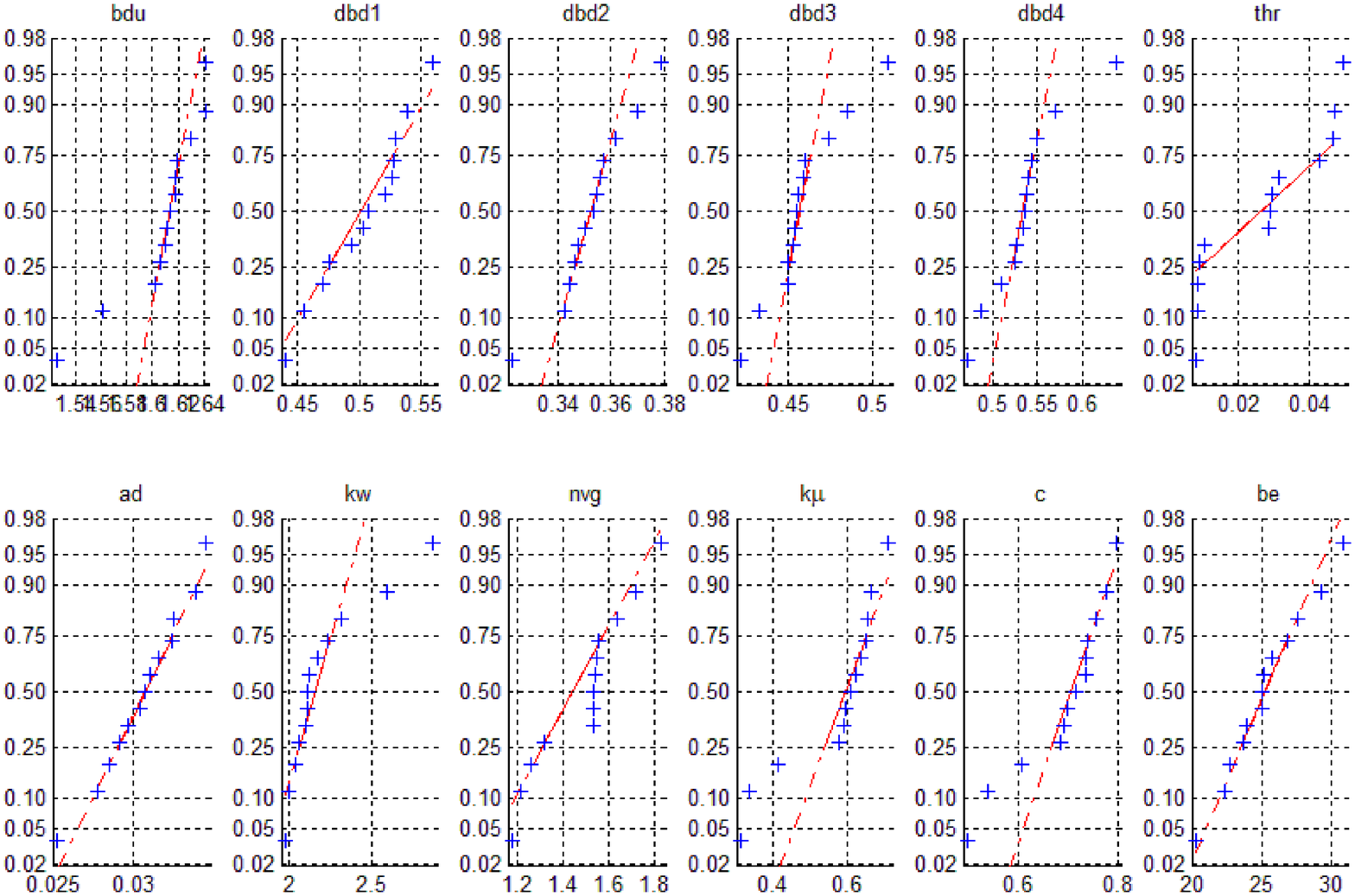}\caption{Parameter values for Plot 2 - irrigated}
\end{figure}

\begin{figure}[h]
\centering{}\includegraphics[width=14cm]{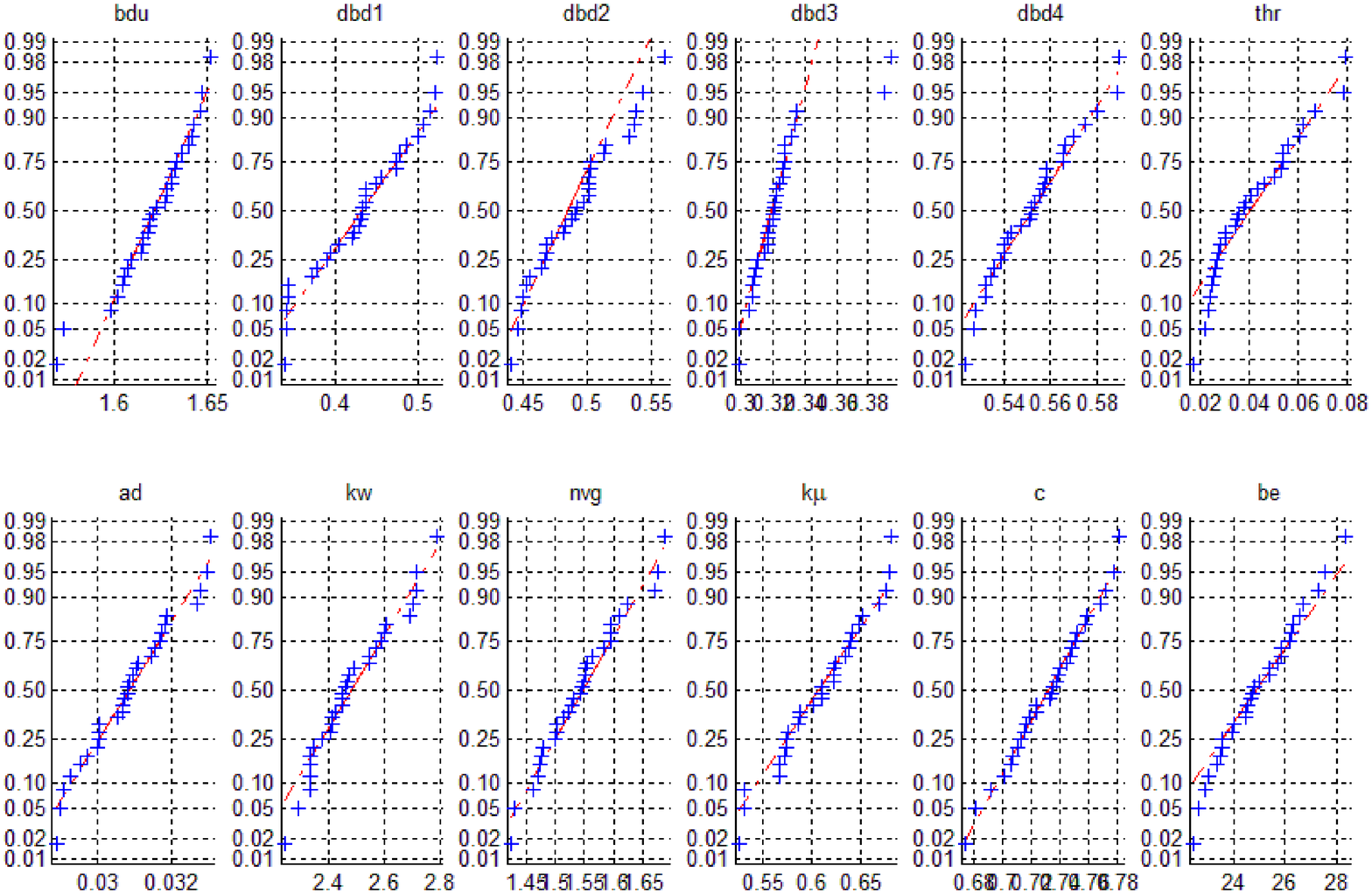}\caption{Parameter values for Plot 3 - irrigated}
\end{figure}

\begin{figure}[h]
\centering{}\includegraphics[width=14cm]{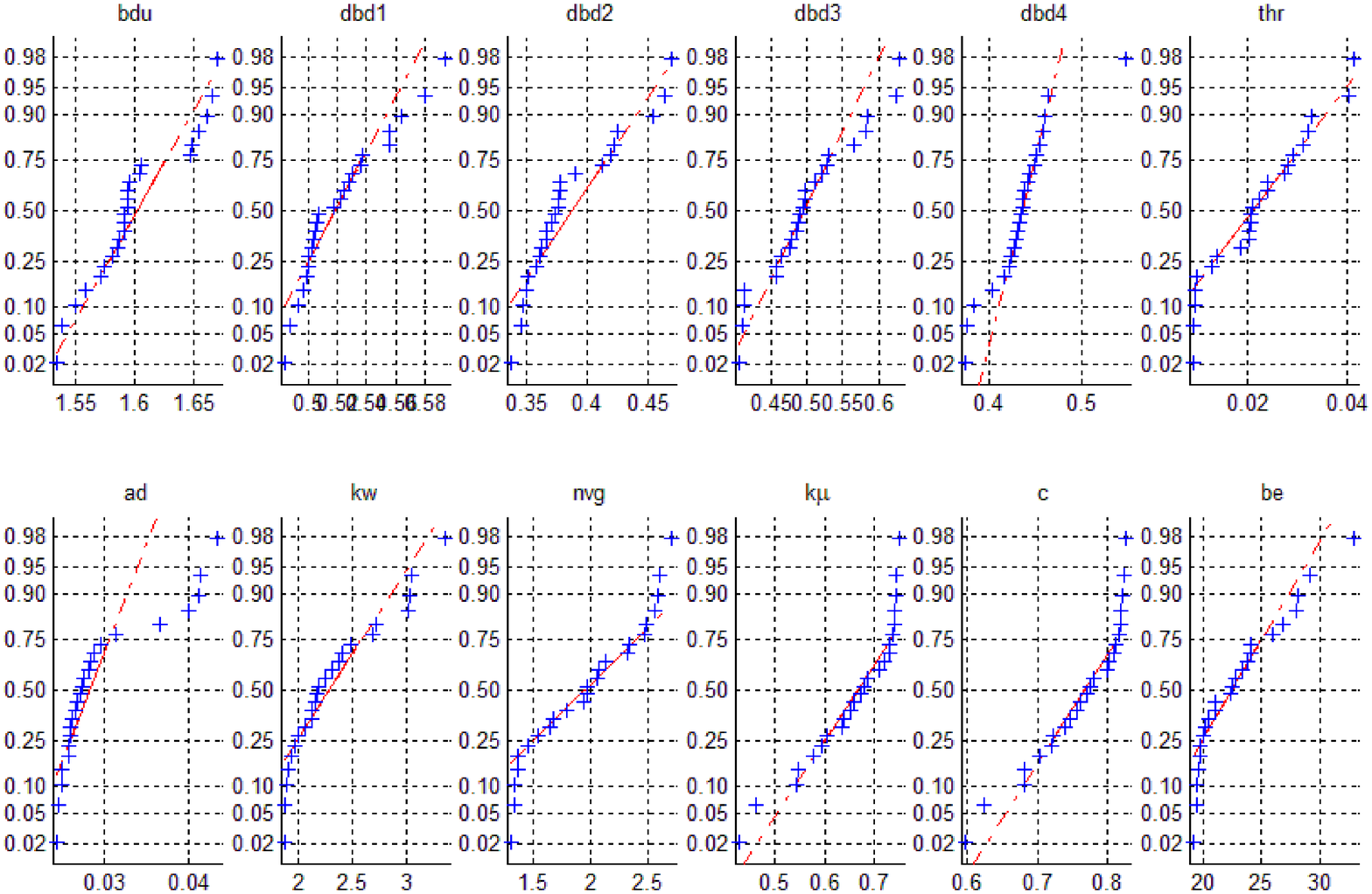}\caption{Parameter values for Plot 3 - rain-fed}
\end{figure}

\begin{figure}[h]
\centering{}\includegraphics[width=14cm]{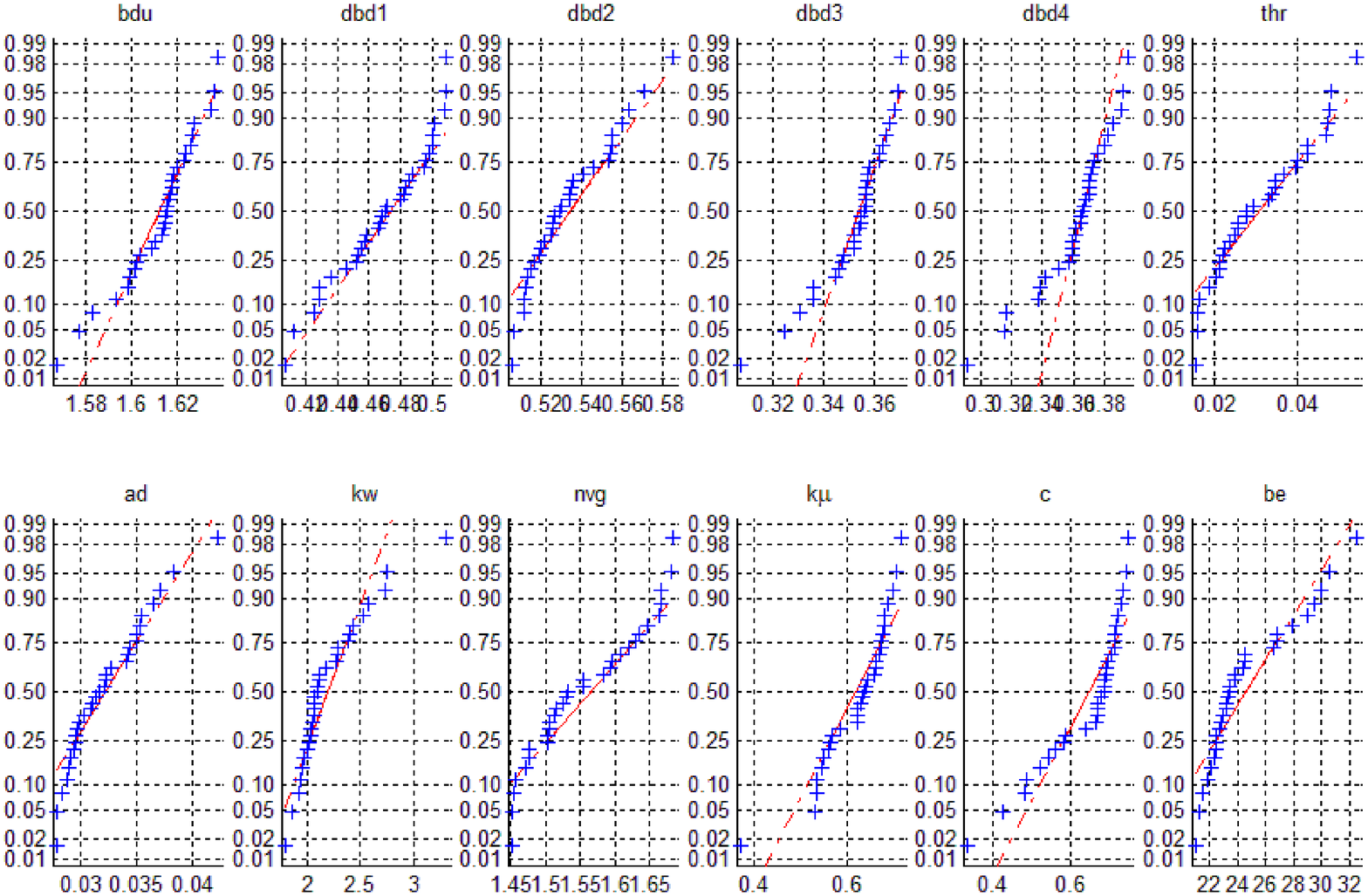}\caption{Parameter values for -Plot 4 - irrigated}
\end{figure}

.
\end{document}